\documentclass[reprint,showpacs,superscriptaddress, english, longbibliography]{revtex4-1}
\usepackage[utf8]{inputenc}
\usepackage{color}
\usepackage{array}
\usepackage{verbatim}
\usepackage{multirow}
\usepackage{amsmath}
\usepackage{amssymb}
\usepackage{dsfont}
\usepackage{graphicx}
\usepackage{esint}
\usepackage[unicode=true,
 bookmarks=true,bookmarksnumbered=true,bookmarksopen=false,
 breaklinks=true,pdfborder={0 0 0},backref=false,colorlinks=true]
 {hyperref}
\usepackage{cleveref}
\usepackage[title]{appendix}
\usepackage{enumitem}

\hypersetup{
citecolor={blue},
urlcolor={blue}
}

\crefname{equation}{Eq.}{Eqs.}
\crefname{figure}{Fig.}{Figs.}
\crefname{section}{Section}{Sections}
\crefname{appendix}{Appendix}{Appendices}
\renewcommand{\paragraph}[1]{\vspace{0.2cm}{\bf \textit{#1}}}
\def\ie{{\it i.e.},\ }
\def\eg{{\it e.g.},\ }
\def\etal{{\it et al.}\ }

\newcommand{\mbf}{\mathbf}
\newcommand{\mcl}{\mathcal}
\newcommand{\mbb}{\mathbb}
\newcommand{\mbs}{\boldsymbol}
\newcommand{\mrm}{\mathrm}
\newcommand{\td}{\widetilde}
\newcommand{\ovl}{\overline}

\def\beq#1\eeq{\begin{equation}#1\end{equation}}
\def\beqs#1\eeqs{\begin{align}#1\end{align}}

\def\pare#1{\left( #1 \right)}
\def\sbrak#1{\left[#1\right]}
\def\bra#1{\left\langle #1 \right|}
\def\ket#1{\left| #1 \right\rangle}

\def\abs#1{\left| #1 \right|}

\def\nono{\nonumber}

\def\vphi{\varphi}
\def\pt{\partial}
\def\pr{\prime}
\def\al{\alpha}
\def\prpr{{\prime\prime}}
\def\tr{\mrm{Tr}}

\def\kk{\mbf{k}}

\def\qq{\mbf{q}} 
\def\hqq{\hat{\qq}}
\def\rr{\mbf{r}}

\def\BB{\mbf{B}}

\def\SS{\mcl{S}}
\def\x{\times}

\def\dn{{\delta n}}
\def\dtn{\td{\dn}}
\def\dbn{\overline{\dn}}
\def\Omg{\mbs{\Omega}} 
\def\omg{\omega}
\def\gm{\gamma}
\def\ee{\epsilon}

\def\sg{\boldsymbol{\sigma}}

\def\vv{\mbf{v}}
\def\hvv{\hat{\mbf{v}}}
\def\odz{{(0)}}
\def\odf{{(1)}}


\begin{document}

\title{Hear the Sound of Weyl Fermions}
\author{Zhida Song}
\affiliation{Department of Physics, Hong Kong University of Science and technology, Clear Water Bay, Kowloon, Hong Kong}
\affiliation{Department of Physics, Princeton University, Princeton, New Jersey 08544, USA}

\author{Xi Dai}
\email{daix@ust.hk}
\affiliation{Department of Physics, Hong Kong University of Science and technology, Clear Water Bay, Kowloon, Hong Kong}

\begin{abstract}
Quasiparticles and collective modes are two fundamental aspects that characterize a quantum matter in addition to its ground state features. 
For example, the low energy physics for Fermi liquid phase in He-III was featured not only by Fermionic quasiparticles near the chemical potential but also by fruitful collective modes in the long-wave limit, including several different sound waves that can propagate through it under different circumstances. On the other hand, it is very difficult for sound waves to be carried by the electron liquid in the ordinary metals, due to the fact that long-range Coulomb interaction among electrons will generate plasmon gap for ordinary electron density oscillation and thus prohibits the propagation of sound waves through it.  In the present paper, we propose a unique type of acoustic collective modes in Weyl semimetals under the magnetic field called chiral zero sound. The chiral zero sound can be stabilized under so-called ``chiral limit", where the intra-valley scattering time is much shorter than the inter-valley one,  and only propagates along an external magnetic field for Weyl semimetals with multiple-pairs of Weyl points. The sound velocity of the chiral zero sound is proportional to the field strength in the weak field limit, whereas it oscillates dramatically in the strong field limit, generating an entirely new mechanism for quantum oscillations through the dynamics of neutral bosonic excitation, which may manifest itself in the thermal conductivity measurements under magnetic field.
\end{abstract}

\maketitle

Topological semimetals are unique metallic systems with a vanishing density of states at the Fermi level \cite{Nielsen1983,Murakami2007,Wan2011,Burkov2011a,Weng2015,Yang2014,Burkov2011b,Kim2015,Yu2015,Fang2015a,Soluyanov2015}. Among different topological semimetals, the Weyl semimetal \cite{Murakami2007,Wan2011,Burkov2011a,Weng2015} is the most robust one, because it requires no particular crystalline symmetry to protect it. The low energy quasiparticle structure of a Weyl semimetal usually contains several pairs of Weyl points (WPs), isolated crossing points in 3D momentum space formed by energy bands without degeneracy. Near each WP, the surrounding quasiparticles can be well described by the Weyl equation proposed by H.E. Weyl 90 years ago in the context of particle physics \cite{Weyl1929}. The WP provides not only the linear energy dispersion around it but more importantly the ``monopole" structure in the Berry's curvature, which makes the dynamics of these Weyl quasiparticles completely different with free electrons in ordinary metals or semiconductors and leads to many exotic properties of the Weyl semimetal, \ie the Fermi arc behavior \cite{Weng2015,Lv2015a,Lv2015b,Xu2015a,Xu2015b,Huang2015,Andolina2018} on the surface and the negative magneto-resistance \cite{Xiong2015,Huang2015b,Li2015,Zhang2016,Li2016} caused by the chiral anomaly \cite{Nielsen1983,Son2012,Stephanov2012,Burkov2014,Gorbar2014}. 

So far the Weyl semimetal is considered as a new topological state in condensed matter physics only because of its unique quasiparticle dynamics, which manifests itself in various transport experiments \cite{Xiong2015,Huang2015b,Li2015,Zhang2016,Li2016}. 
On the other hand, the unique collective modes are another type of features to characterize a new state of matter, which are yet to be revealed for Weyl semimetal systems \cite{Liu2013,Lv2013,Burkov2014b,Stephanov2015,Zhou2015,Pellegrino2015,Song2016,Rinkel2017,Gorbar2017,Andolina2018}. The most common collective mode in a liquid system is sound, which usually requires collisions to propagate. For a neutral Fermi liquid such as He-III \cite{Landau1957,Abrikosov1959,Leggett1975,Volovik2003}, the ordinary sound can only exist when $\omega\tau\ll1$, where $\tau$ is the lifetime of the quasiparticles. For a clean system, the low energy quasiparticle lifetime approaches infinity with reducing temperature, which prohibits the existence of the normal sound modes at enough low temperature when $\omega\tau\gg1$. 
However, there is a completely different type of sound that emerges in the above ``collision-less region" called zero sound, which is purely generated by the quantum mechanical many-body dynamics under the clean limit \cite{Landau1957,Abrikosov1975,Lifshitz2013,Pines1994,Ho1999}. 
In a typical Fermi liquid system, zero sound can be simply viewed as the deformation of the Fermi surface that oscillates and propagates in the system with the ``restoration force" provided by the residual interaction among the quasiparticles around the Fermi surface. 
Like other types of elementary excitations in condensed matters, the form functions of zero sound modes carry the irreducible representations (irreps) of the symmetry group of the particular system. 
For an electron liquid in a normal metal, the density oscillation corresponding to the trivial representation is always governed by the long-range Coulomb interaction and becomes the well-known plasmon excitation with a finite gap in the longwave limit. 
Thus zero sound modes can only exist in high multipolar channels ascribing to the non-trivial representation of the symmetry group, within which the residual interactions among the quasiparticles is positive definite. 
The above condition requires a strong and anisotropic residual interaction in solids, which is difficult to be realized in normal metals.

One of the exotic phenomena of Weyl semimetal is the chiral magnetic effect (CME) \cite{Fukushima2008,Son2012,Zyuzin2012,Vazifeh2013,Chen2013}, where each valley will contribute a charge current under the external magnetic field.
The ``anomalous current'' contributed by CME from a single WP valley with positive (negative) chirality is always parallel (anti-parallel) to the field direction with its amplitude being proportional to the particle number of that particular valley.
To be specific, the ``anomalous current'' contributed by the $\nu$-th valley through CME is $\mbf{j}^a_\nu = e\BB \chi_\nu/(4\pi^2) (\mu_\nu-\mu)$, where $\chi_u$ and $\mu_\nu$ denote the chirality and the imbalanced chemical potential of the $\nu$-th WP, respectively, $\mu$ is the chemical potential at equilibrium, and $e=\mp |e|$ is the charge of electron-like (hole-like) quasiparticle, and $\BB$ is the magnetic field.
The above CME immediately causes an interesting consequence, the particle number imbalance among different valleys will induce particle transport and thus make it possible to form coherent oscillation of the valley particle numbers over space and time, which is a completely new type of collective mode induced by CME.

On the other hand, the most common collective modes in a charged Fermi liquid system are plasmons, and for a Weyl semimetal under a magnetic field, they are such collective modes where the oscillations of the valley particle numbers cannot cancel each other and generate net charge density oscillation in real space \cite{Liu2013,Lv2013,Burkov2014b,Zhou2015,Stephanov2015,Pellegrino2015,Song2016,Rinkel2017,Gorbar2017}. 
Since these modes are coupled to the CME current, the plasmon frequencies significantly depend on the magnetic field \cite{Gorbar2017}.  
Following Ref. \cite{Gorbar2017}, in this paper we call them ``chiral plasmons'' (CPs).
In general, each of the CP modes forms a trivial (identity) irrep of the symmetry group. 
As discussed in detail below, among all the CPs there are only two branches are fully gapped (with opposite frequencies), whereas the other branches are gapless.
For a simplest Weyl semimetal with only a single pair of WPs, the little group at finite wave vector $\qq$ contains only identity operator under magnetic filed, indicating that all the electronic collective modes propagating with wave vector $\qq$ will generally cause net charge density oscillation and thus belong to different branches of the CP modes.

\begin{figure*}
\begin{centering}
\includegraphics[width=0.8\linewidth]{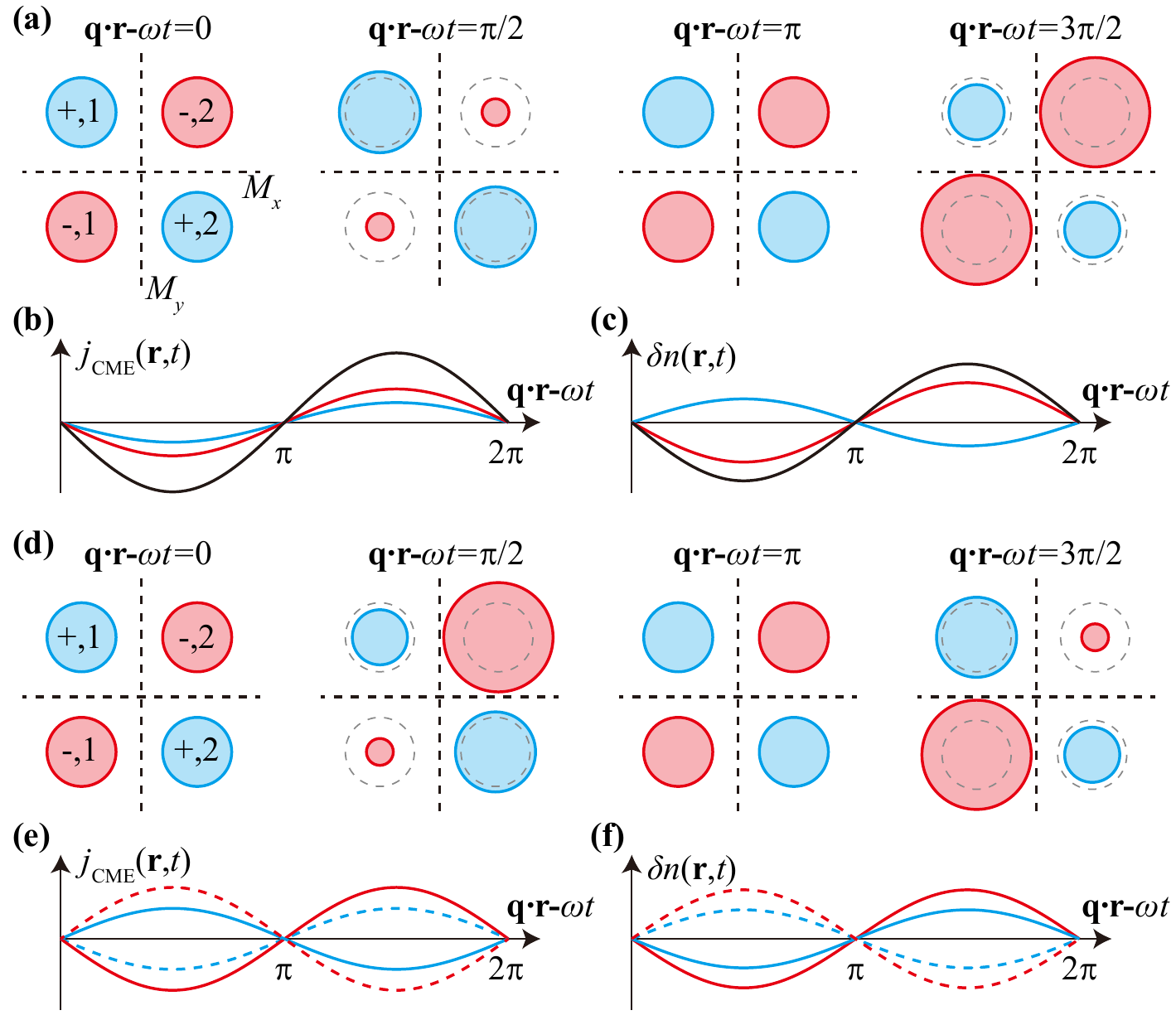}
\par\end{centering}
\protect\caption{Chiral zero sound and chiral plasmon modes in the minimal model with 4 Weyl points. The symmetry group of the model is $C_{2v}$, consisting of a 2-fold rotation axis in $z$-direction, and two mirror planes in $xz$-plane and $yz$-plane, respectively. In (a) and (d), dashed lines represent the two mirrors, the colored discs represent the Fermi surfaces around WPs, and the dashed grey circles represent the Fermi surfaces in equilibrium. 
Here the four WPs are labeled by $(s,a)$, with $s=\pm$ the chirality and $a=1,2$ the sub-valley index.
(a) The volume of Fermi surfaces as functions of space and time in the chiral plasmon mode, where $(s,1)$ and $(s,2)$ are always in phase, making the mode even under $C_2$ rotation.
(b)-(c) The chiral magnetic currents and quasiparticle densities as functions of space and time in the chiral plasmon mode. 
Here the red and blue lines represent the contributions from the $(-,1)$ ($(-,2)$) and $(+,1)$ ($(+,2)$) Fermi surfaces, respectively, and the black lines represent the net current and density.
(d) The volume of Fermi surfaces in the chiral zero sound mode, where $(s,1)$ and $(s,2)$ are always out of phase, making the mode odd under $C_2$ rotation.
(e)-(f) The chiral magnetic currents and quasiparticle densities as functions of space and time in the chiral zero sound mode.
The contributions from $(-,1)$, $(+,1)$, $(-,2)$, and $(+,2)$ Fermi surfaces are represented by the red solid, blue solid, red dashed, and blue dashed lines,  respectively.
In the chiral zero sound mode, both the net current and the net density vanish.
\label{fig:C2v}}
\end{figure*}

The situation becomes completely different for a Weyl semimetal with multiple pairs of WPs. Now we can have collective ``breathing modes" of Fermi surfaces in different WP valleys so that they oscillate in an anti-phase way and cancel out the net charge oscillation exactly, as illustrated schematically in \cref{fig:C2v}(d) for two pairs of WPs. 
Since these anti-phase modes don't cause any net charge current, the collective oscillations of the valley charge and valley current will be completely decoupled from the plasmon modes, and their dispersion relation remains gapless and linear in the longwave limit, which are called “chiral zero sound” (CZS) in this paper. 
As introduced in more detail below, the CZS modes carry the non-trivial irreps of the corresponding little group, with which we can figure out how many CZS modes can exist with the magnetic field being applied in some particular crystal directions.

In order to clearly describe the physical process in Weyl semimetal systems, we divide the charge current contributed by the $\nu$-th WP valley $\bold j_\nu$ into two 
parts, the ``anomalous current" $\bold j_{\nu}^a$ caused by the change of the valley particle number through the CME and the ``normal current"
$\bold j_{\nu}^n$ caused by the deformation of the Fermi surface in the $\nu$-th valley.
For the general situation, the two types of the currents are coupled together and contribute jointly to both of the CP and CME modes. 
However, in the present paper we consider a specific limit, where only the anomalous current can survive, and both the CP and CZS are purely contributed by the CME. Such a limit is proposed previously by D. Son \etal \cite{Son2012}, requiring the intra-valley relaxation time to be much shorter than the inter-valley one, which guarantees the intra-valley relaxation process is fast enough so that any deformation of the Fermi surface from its equilibrium shape can be neglected. 
In the following, we will call this limit as “chiral limit” and mainly discuss the physics of CZS under it.  


Let us first introduce the Boltzmann's equation in the chiral limit.
The Boltzmann's method is valid only in the semi-classical limit, where $\omg_B \tau\ll 1$, $\omg_B\ll \mu$, and $\mu \tau \gg 1$.
Here $\omg_B = v_F\sqrt{eB}$ is the magnetic frequency,  $v_F$ is the Fermi velocity, and $\tau$ is the quasiparticle lifetime. 
(In this paper we set $\hbar=1$, and the energy of WP as $0$.)
In the semi-classical limit, the level smearing caused by finite quasiparticle lifetime is much larger than the Landau-level splitting but is much smaller than the chemical potential, hence the Landau-level quantization can be ignored and the Fermi surface remains well defined.
Therefore, in the semi-classical limit the collective dynamics of a Fermi liquid system can be described by the quasiparticle distribution function $n_{\nu} (\kk,\rr, t)$ through the following Boltzmann equation, where $\nu$ is the valley index, $\kk$ is the momentum, and $\rr$ is the position of the quasiparticle.
\beqs
\frac{d\dn_\nu(\kk,\rr,t)}{dt} =&  \pare{-\dot{\rr}_\nu \cdot \pt_\rr- \dot{\kk}_\nu \cdot \pt_\kk}\dn_\nu(\kk,\rr,t) \nono\\ +& \SS[\dn_\nu(\kk,\rr,t)], \label{eq:Boltzmann-text}
\eeqs
where the first term and second term on the r.h.s. describe the drifting motion and  the scattering process, respectively. 
(An explicit derivation of this equation is given in \cref{app:Boltzmann}).
The time derivative $\dot{\rr}_\nu$ and $\dot{\kk}_\nu$ are given by the equations of motion of the quasiparticles.
In presence of external field and Berry's curvature $\Omg_\nu(\kk)$, they can be written as \cite{Culcer2005,Xiao2010}
\begin{equation}
\gm_\nu(\kk) \dot{\kk}_\nu  = - \pt_{\rr} \ee_{\nu} + e \vv_\nu(\kk) \times \BB + e (\pt_{\rr}  \ee_{\nu}\cdot \BB) \Omg(\kk), \label{eq:EOM-k-text}
\end{equation}
\begin{equation}
\gm_\nu(\kk) \dot{\rr}_\nu = \vv_\nu(\kk) + \pt_{\rr} \ee_{\nu}\times\Omg_\nu(\kk) - e (\Omg_\nu (\kk) \cdot \vv_\nu(\kk)) \BB,\label{eq:EOM-r-text}
\end{equation}
where $\ee_\nu(\kk)$ is the quasiparticle energy, $\vv_\nu(\kk) = \pt_\kk \ee_\nu(\kk)$ is the quasiparticle velocity, and $\gm_\nu(\kk,\BB)=1-e\BB\cdot \Omg_\nu(\kk) $ is the phase-space volume corretion due to the presence of Berry's curvature \cite{Xiao2005}.
We emphasize that $\ee_\nu(\kk)$ is not the bare band energy but the renormalized quasiparticle energy due to the presence of collective mode.
To obtain $\ee_\nu(\kk)$, we first write the total energy in terms of $n_{\nu} (\kk,\rr, t)$
\begin{widetext}
\beqs
E_{\mathrm{total}} (t) &= E_\mrm{total}^0 +  \sum_{\nu}\int {d^3 \rr}{\int \frac{d^3\kk}{(2\pi)^3}} \gm_\nu(\kk,\BB) \ee_{\nu}^0 (\kk) \dn_{\nu} (\kk, \rr, t) + \int \int d^3\rr d^3\rr^{\prime} \frac{e^2}{\ee_0 | \rr-\rr^\pr |} \dn (\rr, t) \dn  (\rr^\pr, t) \nono\\
& + \sum_{\nu \nu^\pr} \int d^3\rr \int \frac{d^3\kk}{(2\pi)^3} \gm_\nu(\kk,\BB) \int \frac{d^3\kk^\pr}{(2\pi)^3} \gm_{\nu^\pr}(\kk^\pr,\BB) f_{\nu, \nu^\pr} \dn_{\nu} (\kk, \rr, t) \dn_{\nu^\pr} (\kk^\pr, \rr, t), \label{eq:Etot-text}
\eeqs
\end{widetext}
where the second and third terms are the long-range Coulomb interaction and the residual short-range interaction between the quasiparticles, respectively \cite{Silin1958}. 
Here $\ee_{\nu}^0 (\kk)$ is the bare energy dispersion for the quasiparticle, $\dn_\nu(\kk,\rr,t)=n(\kk,\rr,t)-n_F(\ee^0_\nu(\kk)-\mu)$ is the deviation from Fermi-Dirac distribution, and $\dn (\mathbf{r}, t) = \sum_{\nu} \int \frac{d^3\kk}{(2\pi)^3} \gm_\nu(\kk,\BB) n_{\nu} (\kk, \rr, t)$ is the net charge density at position $\mathbf{r}$.
In general, the short-range interaction matrix $f_{\nu,\nu^\pr}$ in the above equation should have the full momentum dependence and be written as $f_{\nu,\nu^\pr}(\kk,\kk^\pr)$ \cite{Silin1958}.
However, here we consider the case where the Fermi surfaces are small enough such that the $\kk$-dependence in $f_{\nu,\nu^\pr}(\kk,\kk^\pr)$ can be omitted. 
Then the renormalized quasiparticle energy is given by the functional derivative of the total energy as $\ee_{\nu}(\kk,\rr,t) = \delta E_\mrm{total}/ \dn_\nu(\kk,\rr,t) $.
An elaborate study of collective modes in Weyl systems with only one pair of WPs using the Boltzmann's equation can be found in Ref.~\cite{Stephanov2015}.

To introduce the chiral limit, we decompose $\dn(\kk,\rr,t)$ into two parts: the part that keeps the quasiparticle number in each valley unchanged, $\dtn_\nu(\kk,\rr,t)$, and the part that changes valley quasiparticle numbers, $\dbn_\nu(\kk,\rr,t)$.
In the following we refer to $\dtn_\nu(\kk,\rr,t)$ as the Fermi surface degree of freedom, and $\dbn_\nu(\kk,\rr,t)$ as the valley degree of freedom.
Since the intra-valley scattering preserves the quasiparticle number in each valley, $\dbn_\nu(\kk,\rr,t)$ can be relaxed only through the inter-valley scattering.
On the other hand, $\dtn_\nu(\kk,\rr,t)$ can be relaxed through both the inter- and intra-valley scattering processes.
Therefore, the relaxation time of $\dbn_\nu(\kk,\rr,t)$ is always longer than the the relation time of $\dtn_\nu(\kk,\rr,t)$.
We can approximate the scattering term as
\beq
\SS[\dn_\nu(\rr,\kk,t)] = -\frac{\dtn_\nu(\rr,\kk,t)}{\tau_0} -\frac{\dbn_\nu(\rr,\kk,t)}{\tau_\mrm{v}}. \label{eq:scat-text}
\eeq
As proved in \cref{app:tau_0}, for the simplest case where both the inter- and intra-valley scattering cross sections are constants (without $\mbf{k}$-dependence) \cref{eq:scat-text} is almost exact and the valley degree of freedom has the form $\dbn_\nu(\kk,\rr,t) \propto \dn_\nu(\rr,t) \delta(\ee^0_\nu(\kk)-\mu)$. 
Such a $\kk$-independent scattering cross section is a good approximation for small Fermi surface.
Now we argue that in the chiral limit, where $\tau_0\ll \tau_\mrm{v}$, the Fermi surface degrees and the valley degrees are decoupled and the collective modes are purely contributed by the valley degrees.
To zeroth order of $\tau_0$, nonzero $\dtn_\nu(\kk,\rr,t)$ would be relaxed to zero in an infinitely short time, hence the Fermi surface degrees are always in equilibrium, \ie $\dtn_\nu(\kk,\rr,t)=0$.
Therefore, to obtain the dynamic equation in the chiral limit, we can simply assume $\dn_\nu(\kk,\rr,t) = \dbn_\nu(\kk,\rr,t)$.
Here we take the trial solution as $ \dbn_\nu(\kk,\rr,t)= c_\nu e^{i\qq(\cdot\rr-\omg t)} \delta(\ee^0_\nu(\kk)-\mu)$, where $\qq$ and $\omg$ are the wavevector and frequency of the corresponding collective mode, respectively.
By substituting this trial solution and \cref{eq:EOM-k-text,eq:EOM-r-text} into \cref{eq:Boltzmann-text}, we obtain the following dynamic equation
\beqs
\pare{\omg+\frac{i}{\tau_\mrm{v}}} \eta_\nu =& \frac{e (\qq\cdot\BB)\chi_\nu }{4\pi^2 \beta_\nu(\BB)} \eta_\nu  \nono\\ 
+& \frac{e (\qq\cdot\BB) \chi_\nu}{4\pi^2} \sum_{\nu^\pr} \pare{ f_{\nu,\nu^\pr}  + \frac{e^2}{\ee_0\qq^2} } \eta_{\nu^\pr}, \label{eq:Boltzmann-Son-text}
\eeqs
where $\beta_\nu(\BB)$ is the bare compressibility of the $\nu$-th valley, $\chi_\nu=\pm 1$ is the chirality, and $\eta_\nu = \beta_\nu(\BB) c_\nu$ is the imbalanced quasiparticle particle number (per unit volume) for the $\nu$-th valley.
At zero temperature the bare compressibility is nothing but the density of states at the Fermi level.
In the semiclassical region there are $\beta_\nu(0) \sim \mu^2/v_F^3$ and $\beta_\nu(\BB)- \beta_\nu(0) \sim {\omg_B^2/v_F^3}$.
In the semi-classical limit, $\omega_B\ll\mu$, we have $\beta_\mu(B)\approx \beta_\mu(0)$.

\cref{eq:Boltzmann-Son-text} is the key equation of this paper, which directly leads  to both CP and CZS solutions. 
We put the rigorous derivation in the \cref{app:Chiral} and only give a brief introduction here in the main text. 
We can interprate \cref{eq:Boltzmann-Son-text} as the continuity equation for the quasiparticle number in the $\nu$-th valley under the chiral limit, \ie $\pt_t \eta_\nu + \nabla\cdot \mbf{j}_\nu^a = 0$, where $\mbf{j}_\nu^a$ is the CME current contributed from the $\nu$-th valley.
For simplicity here we set $\tau_\mrm{v}=\infty$.
$i\pt_t \eta_\nu$ gives the l.h.s. and $-i\nabla\cdot \mbf{j}_\nu^a$ gives the r.h.s. of \cref{eq:Boltzmann-Son-text}.
In the chiral limit, each of the Weyl valleys can be described by the Fermi-Dirac distribution functions with time and valley dependent chemical potential $\mu_{\nu}$.
Then the CME current for the $\nu$-th valley $\mbf{j}_\nu^a$ can be simply written as $ \mbf{j}^a_\nu = e\BB\chi_\nu/(4\pi^2) (\mu_\nu-\mu)$, where $\mu$ is the chemical potential in equilibrium.
The above anomalous current $\mbf{j}_{\nu}^a$ is contributed by two effects: the change of quasiparticle number and the modification of the averaged quasiparticle energy in the $\nu$-th valley due to the interaction, which correspond to the two terms in the r.h.s. of \cref{eq:Boltzmann-Son-text} respectively.

In the above analysis, for simplicity, we always neglect the $\kk$-dependence in the form of residual interaction among the quasiparticles, which is a good approximation as long as all the FSs in such Weyl semimetal systems are small enough. 
To generalize our discussion, in \cref{app:k-dep} we prove that even we keep the $\kk$-dependent, the valley degree of freedom $\dbn_\nu(\kk)$ is still well defined and free of intra-valley scattering.
But its form will be modified.
Furthermore, under the chiral limit, the dynamic equation is still given by \cref{eq:Boltzmann-Son-text}, except that $f_{\nu,\nu^\pr}$ has to be understood as the ``$\kk$-averaged'' interaction obtained from  $f_{\nu,\nu^\pr}(\kk,\kk^\pr)$.
Please see \cref{app:k-dep} for more details.

To understand more about the chiral limit, we need to find the upper bound of $\tau_0$ below which the zeroth order discussion is valid.
In \cref{app:tau_0} the effect of finite $\tau_0$ is dealt with the standard second-order perturbation theory. 
Here we only describe the main conclusion: finite $\tau_0$ will introduce an effective damping term $\sim \tau_0 v_F^2 \qq^2$ for the collective modes.
In order to stabilize the collective modes, the Hermitian part of \cref{eq:Boltzmann-Son-text} must be larger than the non-Hermitian part,
or, equivalently, the eigen-frequency should be much larger than the damping rate.
Since the gapped CPs are coupled to the Coulomb interaction, which dominates \cref{eq:Boltzmann-Son-text} in the longwave limit, the conditions for the gapped CPs to be stable are (i) $\frac{1}{\tau_\mrm{v}}\ll \frac{e^3 (\qq\cdot\BB) }{\ee_0 \qq^2}$ (ii) $\tau_0 v_F^2 \qq^2 \ll \frac{e^3 (\qq\cdot\BB) }{\ee_0 \qq^2}$.
These two conditions are automatically satisfied in the longwave limit and hence the gapped CPs are always stable against $\tau_0$. 
On the other hand, since the CZSs and gapless CPs are decoupled from the Coulomb interaction, as shown in the model below and proved generically in \cref{app:Chiral}, the conditions for CZSs and gapless CPs to be stable are (i) $\frac{1}{\tau_\mrm{v}}\ll \frac{e (\qq\cdot\BB) (1+\beta(\BB)f)}{\beta(\BB)}$ and (ii) $\tau_0 v_F^2 \qq^2 \ll \frac{e (\qq\cdot\BB)(1+\beta(\BB)f)}{\beta(\BB)}$.
These two conditions can be satisfied at some $\qq$ only if 
\beq
\frac{\tau_0}{\tau_\mrm{v}} \ll \frac{e^2\BB^2(1+\beta(\BB)f)^2}{v_F^2\beta^2(\BB)}\sim \frac{\omg_B^4}{\mu^4} \pare{1+\frac{\mu^2f}{2\pi^2 v_F^3}}^2. \label{eq:tau0-bound}
\eeq
For simplicity, here we assume isotropic Fermi surfaces such that $\beta(\BB) = \frac{\mu^2}{2\pi^2 v_F^3}$.
Thus the upper bound of $\tau_0$ below which \cref{eq:Boltzmann-Son-text} is valid is given by \cref{eq:tau0-bound}.

Now let us analyze the (magnetic) point symmetry group of \cref{eq:Boltzmann-Son-text}.
Since the wavevector $\qq$ enters \cref{eq:Boltzmann-Son-text} only through the $\qq\cdot\BB$ term, the symmetry group of \cref{eq:Boltzmann-Son-text} is much higher than the little group at $\qq$, in fact all the point group operations or combinations of point group operations and the time-reversal that preserve $\qq\cdot\BB$, $\BB$, and $f_{\nu,\nu^\pr}$ will keep \cref{eq:Boltzmann-Son-text} invariant.
We emphasize that $\qq\cdot\BB$ is invariant under proper rotations and the time-reversal, but changes sign under the inversion, and $\BB$ transforms as a vector under proper rotations, keeps invariant under the inversion, but changes sign under the time-reversal.
Therefore only two types of operations can leave \cref{eq:Boltzmann-Son-text} invariant, proper rotations with axis parallel with $\BB$, and time-reversal followed by two-fold proper rotations with axis perpendicular to $\BB$.
In this paper we denote the group consisting of these symmetry operations as $G_c(\BB)$, which is either a magnetic point group or a point group depending on whether or not it contains combinations of point group and the time-reversal operations.
The solutions of \cref{eq:Boltzmann-Son-text} form the representations for the group $G_c(\BB)$, which can be divided into two categories, the trivial and non-trivial irreps. 
It is then easy to see that the CP solutions belong to the trivial irreps and the CZS solutions belong to the non-trivial ones.
To be specific, as proved in \cref{app:Chiral}, the multiplicity of the trivial irrep, or the number of CPs, is given by
\beq
N_\mrm{CP}(\BB) = \sum_\nu^\pr \frac{|G_0||G_c(\BB)\cap G_\nu|}{|G_c(\BB)||G_\nu|}, \label{eq:number-CP-text}
\eeq
and the multiplicity of the nontrivial irreps, or the number of CZSs, is given by
\beq
N_\mrm{CZS}(\BB) = \sum_\nu^\pr { \frac{|G_0|}{|G_\nu|} - \frac{|G_0||G_c(\BB)\cap G_\nu|}{|G_c(\BB)||G_\nu|} }, \label{eq:number-CZS-text}
\eeq
where the summation of $\nu$ will be carried out over all inequivalent WPs. 
(Two WPs are equivalent if they are related by some symmetry operation.) 
$G_0$ is the maximal (magnetic) point group of the (magnetic) space group, $G_\nu$ is the subgroup of $G_0$ that leaves the $\nu$-th WP invariant, and $|G|$ is the number of elements in $G$.
Here we take the Weyl semimetal TaAs \cite{Weng2015} in space group $I4_1md$ (\#109) as an example to show the usage of \cref{eq:number-CP-text,eq:number-CZS-text}.
Since TaAs is time-reversal symmetric and the maximal point group of $I4_1md$ is $C_{4v}$, we obtain $G_0 = C_{4v} + T C_{4v}$, where $T$ represents the time-reversal.
Totally there are 24 different WPs in TaAs, which can be divided into two classes, 8 WPs located at the $k_z=0$ plane and 16 WPs located off the $k_z=0$ plane. 
The WPs within the same class can be related by operations in $G_0$ and are considered to be equivalent from symmetry point of view.
The corresponding little groups that leave the WPs unchanged are $G_1 = \{E\}$ and $G_2 = \{E, TC_2\}$, respectively.
Therefore, from equation (\ref{eq:Boltzmann-Son-text}), there are totally 24 independent variables leading to same number of independent modes.
Assuming the magnetic field is applied along the $C_4$ rotation axis, we obtain $G_c(\BB) = C_4$ and hence $N_\mrm{CP} = 6$ and $N_\mrm{CZS} = 18$.

As discussed above, in the semiclassical region we always have $\beta_\nu(\BB) - \beta_\nu(0) \sim \frac{\omg_B^2}{\mu^2} \beta_\nu(0)$, which is derived in detail in \cref{app:Chiral}.
Thus, to the leading order effect of the magnetic field, we can omit the $\BB$-dependence in $\beta_\nu(\BB)$.
Then \cref{eq:Boltzmann-Son-text} is in first order of $\BB$ and the corresponding symmetry group becomes higher than $G_c(\BB)$.
This higher symmetry group, denoted as $G_c(0)$, consists of all the proper rotations, time-reversal (if present), and time-reversal followed by proper rotations (if present) in the original group.
Thus $G_c(0)$ is nothing but the chiral subgroup of the little group at $\qq=0$.
Therefore, under semiclassical approximation, the number of CPs and CZSs (\cref{eq:number-CP-text,eq:number-CZS-text}) should be calculated with $G_c(0)$ instead of $G_c(\BB)$.


At last, we consider a model Weyl semimetal system with only two pairs of WPs with point group symmetry $C_{2V}$, as illustrated schematically in \cref{fig:C2v}.
For convenience, we split the valley index $\nu$ into a chirality index $s=\pm 1$ and a sub-valley index $a = 1,2$.
Under the $C_2$ rotation, the $(s,1)$ WP and the $(s,2)$ WP transform to each other;
under the $M_x$ mirror, the $(+,a)$ WP and the $(-,a)$ WP transform to each other.
Thus the representation matrices formed by $\eta_\nu$ can be written as $D_{sa;s^\pr a^\pr}(C_2) = \tau^{x}_{a,a^\pr} \sigma^{0}_{s,s^\pr}$ and  $D_{sa;s^\pr a^\pr}(M_x) = \tau^{0}_{a,a^\pr} \sigma^{x}_{s,s^\pr}$, where $\tau^{x,y,z}$ and $\sigma^{x,y,z}$ are Pauli matrices in the chirality space and sub-valley space, respectively, and $\tau^0$ and $\sigma^0$ are 2 by 2 identity matrices.
In the following, we would omit the matrix subscripts for brevity.
Without loss of generality, we choose the form of residual interaction as $f =  f_0 \tau^{0} \sigma^{0} + f_1 \tau^x \sigma^0 + f_2 \tau^0 \sigma^x + f_3 \tau^x \sigma^x $, where we set $f_0\ge f_1 \pm (f_2-f_3)$ and $f_0\ge - f_1 \pm(f_2+f_3)$ to ensure that the interaction is positive semi-definite.
The magnetic field is applied in the $z$-direction.
Applying the representation matrices to \cref{eq:Boltzmann-Son-text}, one can easily verify that the $C_2$ symmetry is kept but the $M_x$ symmetry is broken.
Thus the solutions will form the irreps of $C_2$.
By diagonalizing \cref{eq:Boltzmann-Son-text}, we obtain two branches of CPs
\beqs
\omg^{(1,2)}(\qq) & + \frac{i}{\tau_\mrm{v}}  = \pm \frac{e(q_zB)}{4\pi^2 \beta(B)} \sqrt{\zeta_0^2(\qq)-\zeta_1^2(\qq)}, \label{eq:CP-text}
\eeqs
where $\zeta_0(\qq) = 1 + \beta(B)(f_0 + f_1 + 2e^2/(\ee_0\qq^2) )$ and  $\zeta_1(\qq) = \beta(B) ( f_2 + f_3 + 2e^2/(\ee_0\qq^2) )$, and two branches of CZSs
\beq
\omg^{(3,4)}(\qq) + \frac{i}{\tau_\mrm{v}} = \pm \frac{e(q_zB)}{4\pi^2 \beta(B)} \sqrt{\xi_0^2-\xi_1^2}, \label{eq:CZS-text}
\eeq
where $\xi_0 =  1+ \beta(B) (f_0 - f_1) $ and $\xi_1 = \beta(B) (f_2 - f_3) $.
In the longwave limit, we have $\zeta_0^2(\qq) -\zeta_1^2(\qq) \approx \frac{4e^2}{\ee_0 \qq^2} \pare{1 + \beta(B)(f_0+f_1-f_2-f_3)} $, so the CP modes are gapped and the plasmon frequency is approximately
\beq
\frac{e^2(q_zB)}{2\pi^2 |\qq|} \sqrt{\frac{1+ \beta(B)\pare{f_0+f_1-f_2-f_3}}{\beta(B)\ee_0}}.
\eeq
On the other hand, the CZS modes have linear dispersions along the magnetic field direction with the sound velocity  
\beq
c(B) = \frac{eB}{4\pi^2\beta(B)}\sqrt{\xi_0^2-\xi_1^2}. \label{eq:cB}
\eeq
Here we give a rough estimation of the sound velocity for a typical Weyl semimetal system. 
For simplicity, we set $f=0$, $B=10\mrm{T}$, $\mu = 30 \mrm{meV}$, $v_F = 2 \mrm{eV \AA} \hbar^{-1}$, then we obtain $c(B) \approx 0.34 \mrm{eV \AA} \hbar^{-1} \approx 5\times 10^4 m/s$.

The eigenvectors of the two CP modes are
\beq
\eta^{(1,2)} = \sbrak{\lambda_{1,2}(\qq),\;\; -1,\;\; \lambda_{1,2}(\qq),\;\;-1}^T,
\eeq
where $\lambda_{1,2}(\qq) = (\zeta_0(\qq)\pm \sqrt{\zeta_0^2(\qq)-\zeta_1^2(\qq)})/\zeta_1(\qq)$; and the eigenvectors of the two CZS modes are
\beq
\eta^{(3,4)} = \sbrak{\lambda_{3,4},\;\; -1,\;\; -\lambda_{3,4},\;\; 1}^T,
\eeq
where $\lambda_{3,4} = (\xi_0\pm\sqrt{\xi_0^2-\xi_1^2})/\xi_1$.
In the above expressions, the bases of the $\eta$ vector are ordered as $(s,a)=(+,1)$, $(-,1)$, $(+,2)$, $(-,2)$.
$\eta^{(1,2)}$ are invariant under $C_2$ and hence form the trivial irrep, whereas $\eta^{(3,4)}$ will change sign under $C_2$ and hence form the nontrivial irrep.
The CP mode $\eta^{(1)}$ and the CZS mode $\eta^{(3)}$ are schematically plotted in \cref{fig:C2v} (a) and (d), respectively.
We can find clearly from \cref{fig:C2v} that the CP is such a mode that the quasiparticle densities with the same chirality oscillate with the same phase, while the quasiparticle densities with the opposite chiralities oscillate with opposite phases. 
Since the CME current from the $\nu$-th valley, $\mbf{j}_\nu$, is proportional to $\chi_\nu \eta_\nu$, a net current oscillation will be generated by the CP mode, which couples to the long-range Coulomb interaction and leads to a finite plasmon frequency in the long wave length limit..
In contrast, in the CZS mode the valley densities with the same chirality oscillate with opposite phases, leading to the exact cancellation of CME currents from differen valleys.
Therefore the CZS mode will be completely decoupled from the charge dynamics and can keep its acoustic nature in the long wave length limit.

\begin{figure}
\begin{centering}
\includegraphics[width=1\linewidth]{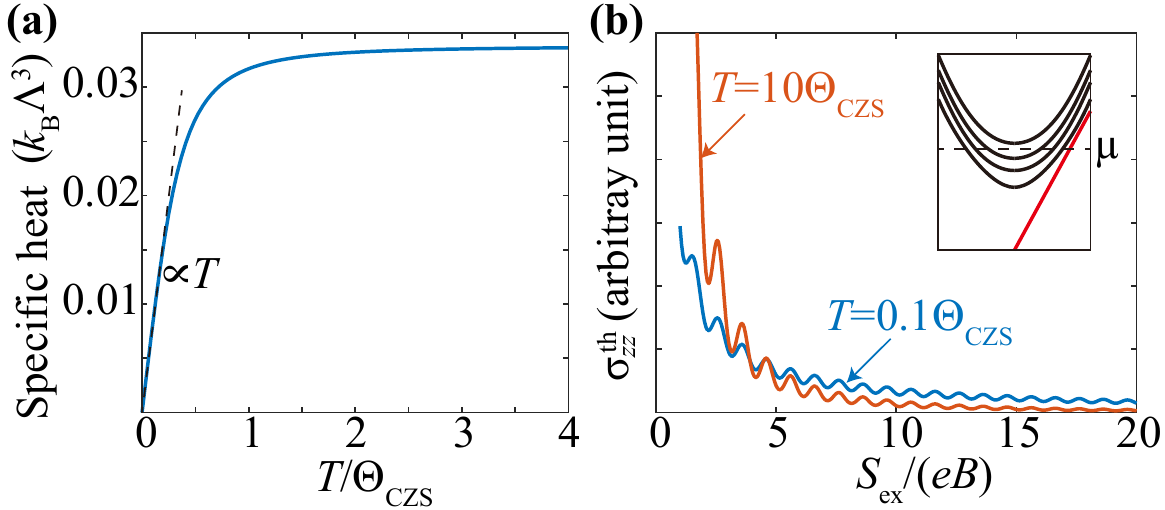}
\par\end{centering}
\protect\caption{ 
(a) The specific heat (per unit volume) in the 4 WPs model is plotted as a function of temperature. 
The specific heat is plotted in the unit of $k_\mrm{B} \Lambda^3$, where $k_\mrm{B}$ is the Boltzmann's constant and $\Lambda$ is the cutoff in momentum integral.
The temperature $T$ is plotted in the unit of the Debye temperature for the CZS mode, $\Theta_\mrm{CZS} = c(B) \Lambda/k_\mrm{B}$, where $c(B)$ is the speed of CZS.
(b) The thermal conductivity in the 4 WPs model is plotted as a function of magnetic field. 
Here $S_{\mrm{ex}}(\mu)$ is the area enclosed by the extreme circle on the Fermi surface that is perpendicular to the magnetic field.
The parameters are set as $\omg_B = 0.2\mu$, $k_\mrm{B}T= 1/(2\tau_0) = 0.001\mu$, and $T/\Theta_\mrm{CZS}=10$ and $0.1$ for the blue line and red line, respectively. 
\label{fig:Oscillation}}
\end{figure}

It is insightful to compare the possibility to have zero sound modes in ordinary metals and Weyl semimetals under magnetic field. The collective modes
for the former have been discussed in detail in Ref. \cite{Pines1994}. Using the 
description developed above, for an ordinary metal, all the collective modes can be derived from the dynamics of Fermi surface degree of freedom 
$\dtn_\nu(\rr,\kk,t)$, which describes the small deviation of the quasiparticle occupation at the Fermi surface. For a system with approximately the 
sphere symmetry, it can be expanded using the sphere harmonics $Y_{l m}(\theta_k,\phi_k)$. Therefore, the longitudinal mode is formed by the proper
linear combination of the sphere harmonics with $m=0$ and becomes the plasmon mode.
The transverse modes are described by the sphere harmonics with  $m\neq 0$. Among them, the channel with $m=\pm 1$ will be absorbed into the Maxwell equation to describe the possible electrical magnetic wave, which contains no solution for frequency below plasmon edge. 
Therefore, the only possible channels to have zero sound modes in an ordinary
metal system are the channels with $\abs{m}\ge 2$ provided that the effective residual interaction in these channels are positively definite to survive the 
Landau damping. 
These conditions are difficult to fulfil and so does zero sound in ordinary metal. Therefore, for Weyl semimetals under magnetic
field, the CME provides a unique mechanism to stabilize the CZS with any form of residual interaction that does not cause instability. 
At least in the chiral limit, the dynamics of CZS only involves the anomalous current but not the normal current, and hence is free of Landau damping.

The existence of CP and CZS in the chiral limit leads to several interesting physical phenomena under the external magnetic field. 
Here we introduce two of them.
The first one is the CZS contribution to the specific heat. 
The CZS modes can be viewed as a set of 1D collective modes dispersing only along the magnetic field. 
As derived in \cref{app:thermal}, the specific heat contributed by the CZS is $\kappa(B,T) \approx k_\mrm{B}^2 T \Lambda^2 /( 6 c(B) )$ for temperature $T\ll \Theta_\mrm{CZS}$, where $\Theta_\mrm{CZS}=c(B) \Lambda/k_\mrm{B}$ is the corresponding Debye temperature for the CZS, $c(B)$ is the sound velocity (\cref{eq:cB}), $\Lambda$ is the momentum space cutoff, and $k_\mrm{B}$ is the Boltzmann's constant.
While, in the high temperature region ($T\gg \Theta_\mrm{CZS}$), the specific heat is $\kappa(B,T) \approx k_\mrm{B} \Lambda^3 /(3\pi^2)$. 
To be specific, in \cref{fig:Oscillation}(a) we plot the specific heat as the function of temperature using some typical parameters for the Weyl semimetal systems.
Although such temperature dependence is similar with the quasiparticle contribution to the specific heat, the two can be distinguished from each other the latter by their different field dependence. 
Another unusual property caused by the CZS is the thermal conductivity.
Since the CZS disperses only along the field direction, the thermal current carried by the CZS modes can only flow along this direction.
As the result, the thermal conductivity tensor contributed by the CZS modes has only one nonzero entry.
As derived in \cref{app:thermal}, if the magnetic field is applied along the $z$-direction, the thermal conductivity is given by $\sigma^\mrm{th}_{ij} = \delta_{i,z} \delta_{j,z} \tau_\mrm{s}(T) c^2(B) \kappa(B,T)$, where $\tau_\mrm{s}(T)$ is the relaxation time for the CZS excitations.
In the weak field and low temperature region ($\tau_0\omg_B\ll 1$, $T \ll \Theta_\mrm{CZS}$), as $\kappa(B,T) \propto T/c(B)$ and $c(B)\propto B$, we obtain $\sigma^\mrm{th}_{zz} \propto TB$.
In the weak field and high temperature region ($\tau_0\omg_B\ll 1$, $T \gg \Theta_\mrm{CZS}$), as $\kappa \sim const.$, we obtain $\sigma^\mrm{th}_{zz} \propto B^2$.

In order to discuss the specific heat and thermal conductivity in the strong field region ($\tau_0\omg_B \gtrsim 1$), we need to re-derive the dynamic equation under the strong field, where the electronic states are already Landau levels.
In this case, since the compressibility oscillates with the field, as a consequence the velocity of the CZS, as well as the thermal conductivity in general, should also oscillate with the field.
Here we only focus on the case $\omg_B\ll\mu$ so that there are still a large number of Landau levels below the chemical potential.
As introduced in \cref{app:strongB}, it turns out that the dynamic equation has the same form of \cref{eq:Boltzmann-Son-text}, except that the field dependence of the compressibility is modified.
As calculated in \cref{app:dos-B}, the compressibility in strong feild can be expressed as $\beta(B) = \beta^{(0)}(B) + \beta^{(1)}(B) + \cdots$, where the $\beta^{(l\ge1)}(B)$ terms oscillate as the $l$-th harmonics  of $1/{ B}$. 
Under the finite temperature, the ratio between the first and zeroth components is approximately
\beq
\frac{\beta^\odf(B)}{\beta^\odz(B)} 
\approx \frac{\omg_B}{\mu} \frac{ \exp\pare{ - \pi \frac{\mu }{\omg_B^2 \tau_0}  } }{ \mrm{sinch}\pare{2\pi^2\frac{\mu k_\mrm{B}T}{\omg_B^2}} }
\cos\pare{\frac{S_{\mrm{ex},\nu}(\mu)}{eB} - \frac{\pi}{4}}, \label{eq:SdH-beta}
\eeq
where $\mrm{sinch}(x) = (e^x-e^{-x})/(2x)$, and $S_{\mrm{ex}}(\mu)$ is the area enclosed by the extreme circle (perpendicular to $\BB$) on the Fermi surface.
Here we have assumed $eB>0$ and $\mu>0$.
Due to \cref{eq:Boltzmann-Son-text,eq:cB}, the oscillation in $\beta(B)$ will lead to the oscillation in the sound velocity of CZS.
Substituting \cref{eq:SdH-beta} to \cref{eq:cB}, we obtain the first order oscillation of the sound velocity as
\beq
c(B) \approx c^{(0)}(B) \pare{  1 - \frac{\xi_0}{\xi_0^2-\xi_1^2} \frac{\beta^{(1)}(B)}{\beta^{(0)}(B)} },
\eeq
where $c^{(0)}(B)$ is the non-oscillating component of the sound velocity.
As both the specific heat and thermal conductivity are functions of sound velocity, the oscillation in velocity leads to the oscillations in specific heat and thermal conductivity as well.
As an example, in \cref{fig:Oscillation}(b) we plot the thermal conductivity as a function of magnetic field.
In normal metals, the thermal conductivity is mainly contributed by electrons and acoustic phonons. 
The phonon part only couples indirectly to the magnetic field and usually doesn’t change much with the field. 
Therefore the part that oscillates with the field is mainly contributed by the free electrons in the normal metal, which satisfies the Wiedemann-Frantz law.
While as we have introduced above, for the Weyl semimetals in the chiral limit,  since the CZS can only propagate along the magnetic field, the thermal conductivity along the field will be contributed by both the CZS and free electrons leading to the dramatic violation of Wiedemann-Frantz law, which is absent for thermal conductivity along the perpendicular direction. 
Early theoretical studies on the electronic contribution to the thermal conductivity in Weyl semimetal without considering the CZS modes obtain the $\BB^2$ dependence for the thermal conductivity under magnetic field \cite{Lundgren2014}, which is quite different with the contribution from the CZS introduced above.
Such a field dependent violation of the Wiedemann-Frantz law had already been seen in the thermal conductivity measurement of TaAs under a magnetic field, indicating the possible contribution from the CZS.
We would note that for realistic systems, which are not deeply in the chiral limit, the CZS will also acquire nonzero velocity along the transverse
direction of the magnetic field as well, which is caused by the accompany normal current during the oscillation. Therefore, the CZS or the gapless CP 
can also contribute to the thermal conductivity along the transverse direction but the effect should be much less by orders than that of the longitudinal direction.

The above mentioned quantum oscillations in specific heat and thermal conductivity can be viewed as strong evidence for the existence of CZS but still indirect. It will be much convincing, if we can also have direct ways to measure it. On this regard, the direct ultrasonic measurement of these materials 
under magnetic field and low temperature may be difficult but worth trying. Another possible experiment is inelastic neutron scattering spectrum. Although 
the corresponding scattering cross section for electrons may be very small, the existence of CZS can still be inferred from the spectrum of certain phonon modes, which have the same symmetry representation with the CZS and can hybridize with it when they intersect each other at some particular wavevector to form the ``polariton mode".

In summary, we have proposed that an exotic collective mode, the chiral zero sound, can exist in a Weyl semimetal under magnetic field with the chiral limit, where the inter-valley scattering time is much longer than the intra-valley one. 
The CZS can propagate along the external magnetic field with its velocity being proportional to the field strength in the weak field limit and oscillating in the strong field. 
The CZS can lead to several interesting phenomena, among which the giant quantum oscillation in thermal conductivity is the most striking and can be viewed as the “smoking gun” evidence for the existence of it.

\textit{Acknowledgement.} Z.S. and X.D. acknowledge  financial  support  from the  Hong  Kong  Research  Grants  Council  (Project No. GRF16300918).  
Z.S. also acknowledges the Department of Energy Grant No. desc-0016239, the National Science Foundation EAGER Grant No. DMR 1643312, Simons Inves-tigator Grants No. 404513, No. ONR N00014-14-1-0330, No. NSF-MRSECDMR DMR 1420541, the Packard Foundation No. 2016-65128, the Schmidt Fund for Development of Majorama Fermions funded by the Eric and Wendy Schmidt Transformative Technology Fund.

\bibliography{ref}

\appendix
\widetext

\section{Boltzmann's equation and collective modes} \label{app:Boltzmann}
Let us first derive the Boltzmann's equation, which applies when the Landau level splitting, \ie $\omg_B = v_F \sqrt{eB}$, is smaller than the imaginary part of the quasiparticle self-energy and the chemical potential, $\mu$.
The semiclassical equations of motion of Weyl fermion are \cite{Culcer2005,Xiao2010}
\beq
\dot{\kk}=-\pt_\rr \ee(\kk,\rr,t)+e\dot{\rr} \x \BB,
\eeq
\beq
\dot{\rr}=\pt_\kk \ee(\kk,\rr,t)-\dot{\kk}\x\Omg\pare{\kk},
\eeq
where $e=- |e|$ ($|e|$) for electron-like (hole-like) quasiparticle, and
\beq
\Omg(\kk) = -i \bra{\pt_\kk u(\kk)} \x \ket{\pt_\kk u(\kk)}
\eeq
is the Berry's curvature.
The decoupled equations are
\beq
\gm(\kk,\BB) \dot{\kk} = -\pt_\rr \ee(\kk,\rr,t)+ e \pt_\kk \ee(\kk,\rr,t)\x\BB + e(\pt_\rr \ee(\kk,\rr,t)\cdot\BB) \Omg(\kk),\label{eq:dotk}
\eeq
\beq
\gm(\kk,\BB) \dot{\rr} = \pt_\kk \ee(\kk,\rr,t) +  \pt_\rr \ee(\kk,\rr,t)\x\Omg(\kk) -e (\pt_\kk\ee(\kk,\rr,t)\cdot\Omg(\kk)) \BB,\label{eq:dotr}
\eeq
where
\beq
\gm(\kk,\BB) = 1-e\BB\cdot\Omg(\kk)
\eeq
is the phase space measure.
Now we denote the distribution function over phase space as $\rho(\kk,\rr,t)$, due to particle number conservation, we have
\beq
\rho(\kk+dt\dot{\kk},\rr+dt\dot{\rr},t+dt)(1+dt\partial_\kk\cdot\dot{\kk}+dt\partial_\rr\cdot\dot{\rr})d^3\kk d^3\rr
=\rho(\kk,\rr,t)d^3\kk d^3\rr, 
\eeq
and hence
\beqs
0 & = \frac{\pt}{\pt t}\rho(\kk,\rr,t) + \sbrak{(\partial_\kk \cdot \dot{\kk}) + \dot{\kk}\cdot\partial_\kk + \pt_\rr\cdot\dot{\rr} + \dot{\rr}\cdot\pt_\rr}\rho(\kk,\rr,t) \nono\\
&= \frac{\pt}{\pt t}\rho(\kk,\rr,t) + \pt_\kk\cdot(\dot{\kk}\rho(\kk,\rr,t)) + \pt_\rr \cdot (\dot{\rr}\rho(\kk,\rr,t)).
\eeqs
Here we have neglected the scattering term in Boltzmann's equation.

From now on, we assume there are a few valleys and label quantities in different valleys with a subscript $\nu$.
For each valley, we introduce a weighted distribution function $n_\nu(\kk,\rr,t) = \rho_\nu(\kk,\rr,t)/\gm(\kk)$, then the multi-valley Boltzmann's equation is given by
\beqs
0 =&\gm_\nu(\kk,\BB) \frac{\pt}{\pt t} n_\nu(\kk,\rr,t) \nono\\
+ & \sbrak{-\pt_\rr \ee_\nu(\kk,\rr,t)+ e \pt_\kk \ee_\nu(\kk,\rr,t)\x\BB + e(\pt_\rr \ee_\nu(\kk,\rr,t)\cdot\BB) \Omg_\nu(\kk)}\cdot\pt_\kk n_\nu(\kk,\rr,t)  \nono\\
+ &  \sbrak{\pt_\kk \ee_\nu(\kk,\rr,t) + \pt_\rr \ee_\nu(\kk,\rr,t)\x\Omg_\nu(\kk) -e (\pt_\kk\ee_\nu(\kk,\rr,t)\cdot\Omg_\nu(\kk)) \BB}\cdot \pt_\rr n_\nu(\kk,\rr,t), \label{eq:n-drift}
\eeqs
where, again, the scattering is neglected.
In derving \cref{eq:n-drift} we have made use of the relations $\pt_\rr \cdot(\gamma_\nu(\kk,\BB) \dot{\rr}_\nu )=0$ and $\pt_\kk \cdot(\gamma_\nu(\kk,\BB) \dot{\kk}_\nu )=0$. 
Due to \cref{eq:dotk,eq:dotr}, this two relations are satisfied as long as (i) $\kk$ is not at the Weyl point, where the semiclassical method does not apply, and (ii) $\pt_\rr \cdot \pt_\kk \ee(\kk,\rr,t)=0$, which is automatically satisfied in our approximation for quasiparticle energy (\cref{eq:QE}).

In presence of collective mode, the single particle energy $\ee_\nu(\kk,\rr,t)$ should be determined self-consistently. 
With quasiparticle excitation, the total energy is a functional of the distribution function \cite{Silin1958}
\beqs
E_{\mathrm{total}} (t) &= E_\mrm{total}^0 +  \sum_{\nu} \int d^3\rr \int \frac{d^3\kk}{(2\pi)^3} \gm_\nu(\kk,\BB) \ee_{\nu}^0 (\kk) \dn_{\nu} (\kk, \rr, t) + \frac12 \int \int d^3\rr d^3\rr^{\prime} \frac{1}{\ee_0 | \rr-\rr^\pr |} \dn (\rr, t) \dn  (\rr^\pr, t) \nono\\
& + \frac12  \sum_{\nu \nu^\pr} \int d^3\rr \int \frac{d^3\kk}{(2\pi)^3} \gm_\nu(\kk,\BB) \int \frac{d^3\kk^\pr}{(2\pi)^3} \gm_{\nu^\pr}(\kk^\pr,\BB) f_{\nu, \nu^\pr}\dn_{\nu} (\kk, \rr, t) \dn_{\nu^\pr} (\kk^\pr, \rr, t), \label{eq:Etot}
\eeqs
where the second and third terms denote the long range Coulomb and residual short range interaction among the quasiparticles around the WPs respectively. 
Here
\beq
\delta n_\nu(\kk,\rr,t) =  n_\nu(\kk,\rr,t) - n_F (\ee^0_\nu(\kk)-\mu)
\eeq 
is the deviation of distribution from equilibrium, $n_F(\ee) = 1/\pare{1+\exp\pare{-\frac{\ee}{k_\mrm{B}T}}}$ is the Fermi-Dirac distribution,  
\beq \dn(\rr, t) = \sum_{\nu} \int \frac{d^3\kk}{(2\pi)^3} \gm_\nu(\kk,\BB) \dn_{\nu} (\kk, \rr, t)\eeq
is the charge density at position $\mathbf{r}$ and time $t$, $f_{\nu,\nu^\pr}$ is a real matrix due to the Hermitian condition of Hamiltonian.
The $\kk$-dependence of $f_{\nu,\nu^\pr}$ is neglected since we consider the case where the Fermi surfaces are very small compared to the Brillouin zone.
The quasiparticle energy can then be derived as the functional derivation of the total energy
\beq
\ee_\nu(\kk,\rr,t) = \ee^0_\nu(\kk) + \sum_{ \nu^\pr} \int \frac{d^3\kk^\pr}{(2\pi)^3}  f_{\nu,\nu^\pr}\gm_{\nu^\pr}(\kk^\pr,\BB)  \delta n_{\nu^\pr}(\kk^\pr,\rr,t) + e \vphi(\rr,t), \label{eq:QE}
\eeq
where $\varphi$ is the scalar potential determined by possion equation
\beq
-\pt_\rr^2 \vphi(\rr,t) = \frac{e}{\ee_0} \dn(\rr,t).
\eeq
Now we assume the deviation from equilibrium takes the form of plane wave
\beq
\dn_\nu(\kk,\rr,t) = \dn_\nu(\kk) e^{i (\qq\cdot\rr-\omg t) }. \label{eq:n_nu}
\eeq
Following this definition, we can rewrite the quasiparticle energy as
\beq
\ee_\nu(\kk,\rr,t) =  \ee^0_\nu(\kk) + \sum_{ \nu^\pr}\int \frac{d^3\kk^\pr}{(2\pi)^3}  \pare{f_{\nu,\nu^\pr} + \frac{e^2}{\ee_0\qq^2}} \gm_{\nu^\pr}(\kk^\pr,\BB)
\dn_{\nu^\pr}(\kk^\pr) e^{i (\qq\cdot\rr-\omg t) }, \label{eq:QE-q}
\eeq
The equation of motion to first order of $\dn_\nu(\kk)$ is given by
\beqs
0 =- & \gm_\nu(\kk,\BB) \omg \dn_\nu(\kk) \nono\\
+ & \pare{\qq\cdot\vv_\nu(\kk) - e\qq\cdot\BB(\vv_\nu(\kk)\cdot\Omg_\nu(\kk))}
\bigg( \dn_\nu(\kk) +  \delta_T(\mu-\ee^0_\nu(\kk)) \sum_{ \nu^\pr}\int \frac{d^3\kk^\pr}{(2\pi)^3} \pare{f_{\nu,\nu^\pr} + \frac{e^2}{\ee_0\qq^2}} \gm_{\nu^\pr}(\kk^\pr,\BB) \dn_{\nu^\pr}(\kk^\pr) \bigg) \nono\\
- & ie (\vv_\nu(\kk) \x \BB)\cdot\pt_\kk \dn_\nu(\kk), \label{eq:Boltzmann0}
\eeqs
where $\vv_\nu(\kk) = \pt_\kk \ee_\nu^0(\kk)$, and $ \delta_T(\ee) = -\pt_\ee n_F(\ee)$.

For convenience, we replace the 3D variable $\kk$ in \cref{eq:Boltzmann0} with an energy $\ee$ and a 2D wave vector $\sg$ on the energy surface.
The integration over $\kk$ in the $\nu$-th valley can be rewritten as 
\beq
\int d^3\kk = \int_0^\infty d\ee \int_\ee d^2\sg \frac{1 }{|\vv_\nu(\ee,\sg)|},
\eeq
where $\int_\ee$ means $\sg$ takes value on the 2D surface with fixed energy $\ee$.
Apparently, the solution of \cref{eq:Boltzmann0} takes the form 
\beq
\dn_\nu(\kk) = -\pt_\ee n_F(\ee-\mu) \dn_\nu(\sg),
\eeq
where $\sg$ takes value on the Fermi surface.
Integrating the energy, \cref{eq:Boltzmann0} becomes
\beqs
0 = - & \frac{\gm_\nu(\sg,\BB)}{|\vv_\nu(\sg)|}\omg \dn_\nu(\sg) \nono\\
+ & \pare{\qq\cdot\hvv_\nu(\sg) - e\qq\cdot\BB(\hvv_\nu(\sg)\cdot\Omg_\nu(\sg))}
\x \bigg( \dn_\nu(\sg) + \sum_{ \nu^\pr}\int \frac{d^2\sg^\pr}{(2\pi)^3} \pare{f_{\nu,\nu^\pr} + \frac{e^2}{\ee_0\qq^2}} \frac{\gm_{\nu^\pr}(\sg^\pr,\BB)}{|\vv_{\nu^\pr}(\sg^\pr)|} \dn_{\nu^\pr}(\sg^\pr) \bigg) \nono\\
- & ie (\hvv_\nu(\sg) \x \BB)\cdot\pt_\kk \dn_\nu(\sg) , \label{eq:Boltzmann1}
\eeqs
where $\hvv_\nu(\sg) = \vv_\nu(\sg)/|\vv_\nu(\sg)|$, and
\beq
\pt_\kk = \vv_\nu(\kk) \frac{\pt}{\pt \ee} +  \sum_{i=1,2} \pt_\kk \sg_i \frac{\pt}{\pt \sg_i}.
\eeq
In \cref{eq:Boltzmann1} all the quantities are defined on the Fermi surfaces, so we omit the energy dependence of these quantities, \eg $\vv_\nu(\sg)$ is a shorthand of $\vv_\nu(\mu,\sg)$.

\section{The chiral limit} \label{app:Chiral}
We can decompose $\dn$ as a part changing particle number in each valley, $\dbn$, and a part deforming the shape of Fermi surface but preserving particle number in each valley, $\dtn$.
We refer $\dbn$ as the valley degree of freedoms and $\dtn$ as the Fermi surface degree of freedom.
In general case these two degrees of freedoms are strongly coupled.
However, as argued below, in the chiral limit, the dynamic of these two degrees of freedoms are decoupled.
In presence of scattering term, $\dn$ in general damps with time, but the valley degrees and the Fermi surface degrees can have different relaxation time.
We denote the relaxation time of $\dtn$ as $\tau_0$ whereas the relaxation time of $\dbn$ as $\tau_\mrm{v}$.
Then the time derivative term in \cref{eq:Boltzmann1} should be replaced by
\beq
\omg \dn_\nu(\sg) \to 
\pare{\omg+\frac{i}{\tau_\mrm{v}}} \dbn_\nu(\sg) + \pare{\omg+\frac{i}{\tau_0}} \dtn_\nu(\sg).
\eeq
The chiral limit refers to the case that $\tau_0$ is much smaller $\tau_\mrm{v}$, \ie
\beq
\frac{\tau_0}{\tau_\mrm{v}}\ll 1.
\eeq
This limit can be achieved when the intra-valley scattering is much stronger than the inter-valley scattering.
In \cref{app:tau_0} we discuss the relaxation times contributed by impurity scattering.
In the simple case in \cref{app:tau_0}, $\dbn_\nu$ is defined as
\beq
\dbn_\nu  = \frac{1}{\beta_\nu(\BB)} \int \frac{d^2\sg}{(2\pi)^3} \frac{\gm_\nu(\sg,\BB)}{|\vv_\nu(\sg)|} \dn_\nu(\sg),
\eeq
where
\beq
\beta_{\nu}(\BB) = \int \frac{d^2\sg}{(2\pi)^3} \frac{\gm_\nu(\sg,\BB)}{|\vv_\nu(\sg)|} = \frac{d \rho_\nu}{d\mu} \label{eq:rho-B}
\eeq
is the compressibility of the $\nu$-th valley, and $\rho_\nu$ is total particle density of the $\nu$-th valley. 
Then \cref{eq:Boltzmann1} can be rewritten as
\beqs
&  \frac{\gm_\nu(\sg,\BB)}{|\vv_\nu(\sg)|}\pare{\pare{\omg+\frac{i}{\tau_\mrm{v} }} \dbn_\nu+ \pare{\omg+\frac{i}{\tau_0}} \dtn_\nu(\sg)} \nono\\
= & +  \pare{\qq\cdot\hvv_\nu(\sg) - e\qq\cdot\BB(\hvv_\nu(\sg)\cdot\Omg_\nu(\sg))}
\x \bigg( \dn_\nu(\sg) + \sum_{ \nu^\pr} \pare{f_{\nu,\nu^\pr} + \frac{e^2}{\ee_0\qq^2}} \beta_{\nu^\pr}(\BB) \dbn_{\nu^\pr} \bigg) \nono\\
- & ie (\hvv_\nu(\sg) \x \BB)\cdot\pt_\kk \dn_\nu(\sg). \label{eq:Boltzmann2}
\eeqs

In the following, we study the physics in zeroth order of $\tau_0$, and leave the discussion on finite $\tau_0$ effect in \cref{app:tau_0}.
To zeroth order of $\tau_0$, Fermi surface degrees of freedom is always in thermal equilibrium, \ie $\dtn_\nu(\sg)=0$: any deviation from equilibrium will be immediately killed by the strong scattering.
By integrating $\sg$ in \cref{eq:Boltzmann2}, we get a {\it generalized eigenvalue equation}
\beq
\pare{\omg+\frac{i}{\tau_\mrm{v}}} \chi_\nu \eta_\nu = \frac{e (\qq\cdot\BB) }{4\pi^2 \beta_\nu(\BB)} \eta_\nu  + \frac{e}{4\pi^2} (\qq\cdot\BB) \sum_{\nu^\pr} \pare{ f_{\nu,\nu^\pr}  + \frac{e^2}{\ee_0\qq^2} } \eta_{\nu^\pr}. \label{eq:Boltzmann-Son}
\eeq
Here $\chi_\nu = \pm 1$ is the chirality of the $\nu$-th valley, and $\eta_\nu = \beta_\nu(\BB) \dbn_\nu $ is the inequilibrium quasiparticle number in the $\nu$-th valley.
In deriving \cref{eq:Boltzmann-Son}, we have applied
\beq
\int d^2\sg \hvv_\nu(\sg)\cdot\Omg_\nu(\sg) = \int d\mbf{S}\cdot \Omg_\nu(\sg) = -2\pi\chi_\nu. 
\eeq

Now let us discuss the symmetry of \cref{eq:Boltzmann-Son}.
Apparently, \cref{eq:Boltzmann-Son} has a higher symmetry than the little group group of $\qq$: it contains all the symmetries that preserve the chiralities of WPs and the direction of magnetic field.
The direction of $\qq$ is irrelevant to the symmetry.
This is because, in the chiral limit, the electric field, proportional to $\qq$, enters the equation only through the $\qq\cdot\BB$ term, and thus only couples to the chiral degree of freedom.
Therefore, finite $\qq$ only breaks the symmetries changing chiralities. 
In the following, we denote the symmetry group of \cref{eq:Boltzmann-Son} as $G_c(\BB)$. 
We emphasize that some anti-unitary symmetry, like time-reversal followed by a crystalline symmetry, can also keep the chiralities and the magnetic field invariant.
And, since $f_{\nu,\nu^\pr}$ is a real matrix, these anti-unitary symmetries act on \cref{eq:Boltzmann-Son} as unitary operators.
The explicit representation matrix of all these symmetries is given in \cref{eq:rep}.

It should be noticed that, to leading order of magnetic field, \ie setting $\beta_\nu(\BB)=\beta_\nu(0)$, \cref{eq:Boltzmann-Son} even has a symmetry higher than $G_c(\BB)$: the magnetic field enters the equation only through term $\qq\cdot\BB$, thus the direction of magnetic field becomes irrelevant to the symmetry.
We denote this higher symmetry group as $G_c(0)$, which consists of all the symmetries preserves the chiralities of WPs.

Solutions of \cref{eq:Boltzmann-Son} must form irreducible representations (irreps) of $G_c(\BB)$. 
As will be shown in next two sections, the trivial irreps of $G_c(\BB)$ always couple to the charge density oscillation, and thus we call the modes forming trivial irreps as chiral plasmons (CPs).
As will be proved, only two of the CPs are gapped, whereas other CPs are gapless in the longwave limit. 
On the other hand, all the nontrivial irreps are decoupled from density oscillation, so we call them the chiral zero sound (CZS).
Now let us calculate the number of trivial irreps in the solution of \cref{eq:Boltzmann-Son}.
We first consider a set of symmetry related WPs in the inner of Brillouin zone, and one of them has the little group $G_W$.
We denote the maximal (magnetic) point group of the space group as $G_0$, then each symmetry-related WP can be represented by a coset representative of $G_0/G_W$
\beq
G_0 = h_1 G_W + h_2 G_W + \cdots .
\eeq
The representation formed by the valley degrees is given by
\beq
\forall g\in G_0, \quad  D_{h,h^\pr}(g) = \begin{cases}
1 & \qquad \text{if}\ g h^\pr \in hG_W \\
0 & \qquad \text{else}
\end{cases}, \label{eq:rep}
\eeq
\beq
\tr D(g) = \begin{cases}
|G_0|/|G_W| & \qquad g \in G_W \\
0 & \qquad g \notin G_W
\end{cases}.
\eeq
Now we reduce $D$ to irreps of $G_c(\BB)$. 
The number of trivial irrep is given by
\beq
\frac{1}{|G_c(\BB)|} \sum_{g\in G_c(\BB)} \tr D(g)  
= \frac{|G_0||G_c(\BB)\cap G_W|}{|G_c(\BB)||G_W|} 
\eeq
Therefore, for a system with a few set of non-equivalent WPs, the number of CP modes and CZS modes are given by
\beq
N_\mrm{CP}(\BB) = \sum_\nu^\pr \frac{|G_0||G_c(\BB)\cap G_\nu|}{|G_c(\BB)||G_\nu|}, \label{eq:number-CP}
\eeq
and
\beq
N_\mrm{CZS}(\BB) = \sum_\nu^\pr { \frac{|G_0|}{|G_\nu|} - \frac{|G_0||G_c(\BB)\cap G_\nu|}{|G_c(\BB)||G_\nu|} }, \label{eq:number-CZS}
\eeq
respectively.
Here $\nu$ sums over all inequivalent WPs, $G_\nu$ is the little group of the $\nu$-th WP.

\section{Chiral zero sound (CZS)} \label{app:CZS}
If $\eta_\nu$ is not a trivial irrep of $G_c(\BB)$, there must be $\sum_\nu \eta_\nu =0$, implying that it does not cause any charge density oscillation.
Thus for nontrivial irreps we can omit the Coulomb term, and the corresponding modes are the CZSs.
Now let us solve the equation of motion for CZS.
Notice that the matrix in the r.h.s. of \cref{eq:Boltzmann-Son} is real and symmetric, so we diagonalize it as
\beq
f_{\nu,\nu^\pr} + \frac{1}{\beta_\nu(\BB)} \delta_{\nu,\nu^\pr} =  \frac{1}{\ovl{\beta}(\BB)} \sum_a O_{\nu,a} \lambda_a O_{\nu^\pr,a},\label{eq:CZS-matrix}
\eeq
where $\ovl{\beta}(\BB)$ is the averaged $\beta_\nu(\BB)$, $O$ is an orthogonal matrix, and $\lambda_a$'s are dimensionless numbers. 
Applying the transformation
\beq
\eta_\nu = \sum_a O_{\nu,a}\eta_a^\pr,
\eeq
we can rewrite \cref{eq:Boltzmann-Son} as
\beq
\pare{\omg +\frac{i}{\tau_\mrm{v}}} \sum_{\nu,a^\pr} O_{\nu,a} \chi_\nu O_{\nu,a^\pr} \eta_{a^\pr}^\pr = \frac{e}{4\pi^2}(\qq\cdot\BB) \lambda_a \eta_a^\pr.
\eeq
Applying the transformation 
\beq
\Xi_{a,a^\pr} = \sum_\nu \frac{1}{\sqrt{\lambda_a}} O_{\nu,a} \chi_\nu O_{\nu,a^\pr}  \frac{1}{\sqrt{\lambda_{a^\pr}}},\qquad
\eta_a^\prpr = \sqrt{\lambda_a} \eta_a^\pr, \label{eq:CZS-Xi}
\eeq
we get a regular eigenvalue problem
\beq
\Xi \eta^\prpr = \frac{e(\qq\cdot\BB) }{4\pi^2 \ovl{\beta}(\BB) \pare{\omg+\frac{i}{\tau_\mrm{v}}} } \eta^\prpr, 
\eeq
where $\Xi$ has the symmetry of $G_c(\BB)$. 
The dispersion of CZS is given by
\beq
\omg_{\mrm{CZS},n}(\qq) +\frac{i}{\tau_\mrm{v}}= \frac{e (\qq\cdot\BB)}{4\pi^2 \ovl{\beta}(\BB) \xi_n} , \qquad n=1\cdots N_\mrm{CZS}(\BB). \label{eq:omg_CZS}
\eeq
Here $\xi_n$ is the $n$-th eigenvalue of $\Xi$.

It should be noticed that $\Xi$ is real and symmetric (such that $\xi_n$'s are real) only if all $\lambda_a$'s are positive.
Thus the number of CZS modes is given by \cref{eq:number-CZS} only if \cref{eq:CZS-matrix} is positive definite.
Otherwise, only irreps where all $\lambda_a$'s are positive correspond to physically observable modes.
The irreps having negative $\lambda_a$'s in general have complex $\xi$ and so are not stable.

\section{Chiral plasmon (CP)}\label{app:CP}
For the trivial irreps, in general we have $\sum_\nu \eta_\nu \neq 0$. 
Therefore the trivial irreps contribute to density oscillation and thus the Coulomb term must be considered.
However, as $\frac{e^2}{\ee_0 \qq^2}$ is a rank-1 operator, in the longwave limit, there should be only one channel that responses to Coulomb interaction.
To separate this channel, we define the projection operator $P_{\nu,\nu^\pr} = \frac{1}{N_W}$, where $N_W$ is the number of WPs, and divide the terms in the r.h.s. of \cref{eq:Boltzmann-Son} to four components:
\beqs
\sbrak{ \frac{e^2}{\ee_0\qq^2}} + f + \sbrak{\frac{1}{\beta(\BB)} } = & 
P \pare{\sbrak{ \frac{e^2}{\ee_0\qq^2} } + f +  \sbrak{ \frac{1}{\beta(\BB)} } } P + Q\pare{f +  \sbrak{ \frac{1}{\beta(\BB)} } }Q \nono\\
& + P\pare{f + \sbrak{ \frac{1}{\beta(\BB)} } }Q  + Q\pare{f + \sbrak{ \frac{1}{\beta(\BB)} } }P
\eeqs
Here $\sbrak{ \frac{e^2}{\ee_0\qq^2} } $ represents the matrix where every element is $\frac{e^2}{\ee_0\qq^2} $, $\sbrak{ \frac{1}{\beta(\BB)} }$ represents the diagonal matrix $\frac{1}{\beta_\nu(\BB)}\delta_{\nu,\nu^\pr}$, and $Q=\mbb{I}-P$, where $\mbb{I}$ is the identity matrix.
We apply an orthogonal transformation $V=\mathbb{I}+S-S^T$, where $S=PSQ$, to remove the mixing term between $P$ and $Q$ subspaces. 
To second order of $\qq$, we find that
\beq
S = -\frac{\ee_0}{e^2 N_W} \qq^2 P\pare{f+\sbrak{\frac1{\beta(\BB)}}}Q + \mcl{O}(\qq^4),
\eeq
and
\beq
V^T \pare{ \sbrak{ \frac{e^2}{\ee_0\qq^2} } + f + \sbrak{ \frac{1}{\beta(\BB)} } } V = P \pare{\sbrak{ \frac{e^2}{\ee_0\qq^2} } + f +  \sbrak{ \frac{1}{\beta(\BB)} } } P + Q\pare{f +  \sbrak{ \frac{1}{\beta(\BB)} } }Q + \mcl{O}(\qq^2).
\eeq
Using the fact $\sum_{\nu^\prpr} V_{\nu^\prpr,\nu}\chi_{\nu^\prpr} V_{\nu^\prpr,\nu^\pr} = \chi_\nu \delta_{\nu,\nu^\pr} + \mcl{O}(\qq^2)$, we can rewrite \cref{eq:Boltzmann-Son} as
\beq
(\omg + \frac1{\tau_\mrm{v}}) \chi_\nu \eta_\nu =  \sum_{\nu^\pr} \pare{ P \pare{\sbrak{ \frac{e^2}{\ee_0\qq^2} } + f +  \sbrak{ \frac{1}{\beta(\BB)} } } P + Q\pare{f +  \sbrak{ \frac{1}{\beta(\BB)} } }Q }_{\nu,\nu^\pr} \eta_{\nu^\pr} + \mcl{O}(\qq^2).
\eeq

To solve this generalized eigenvalue equation, we apply the technique used in \cref{app:CZS}: diagonalizing the  matrix in the r.h.s. and transforming the equation to a regular eigenvalue problem. 
Let us write the matrix in the write hand side as $\frac1{\ovl{\beta}(\BB)}\sum_a O_{\nu,a} \lambda_a O_{\nu^\pr,a}$.
Applying the transformation 
\beq
\Xi_{a,a^\pr} = \sum_\nu \frac{1}{\sqrt{\lambda_a}} O_{\nu,a} \chi_\nu O_{\nu,a^\pr}  \frac{1}{\sqrt{\lambda_{a^\pr}}},\qquad
\eta_a^\prpr = \sqrt{\lambda_a} \sum_\nu O_{\nu,a} \eta_\nu, \label{eq:CP-Xi}
\eeq
we get a regular eigenvalue problem
\beq
\Xi \eta^\prpr = \frac{e(\qq\cdot\BB) }{4\pi^2 \ovl{\beta}(\BB) \pare{\omg+\frac{i}{\tau_\mrm{v} }} } \eta^\prpr.
\eeq
The frequencies of CP modes are then given by
\beq
\omg_{\mrm{CP},n} + \frac{i}{\tau_\mrm{v}} = \frac{e(\qq\cdot\BB)}{4\pi^2\ovl{\beta}(\BB)\xi_n(\qq)},  \label{eq:omg_CP}
\eeq
where $\xi_n$'s are eigenvalues of $\Xi$.
Now let us analyse the spectrum of $\Xi$.
For convenience, we set $O_{\nu,1}$ as the eigenvector of $P$, so the corresponding eigenvalue is 
\beq
\lambda_1 = \frac{e^2 N_W}{\ee_0 \qq^2} \ovl{\beta}(\BB) + \frac{1}{N_W} \sum_{\nu,\nu^\pr} \pare{ f_{\nu,\nu^\pr} \ovl{\beta}(\BB) + \frac{\ovl{\beta}(\BB)}{\beta_\nu(\BB)}\delta_{\nu,\nu^\pr} } + \mcl{O}(\qq^2),
\eeq
which is singular in the limit $\qq\to 0$.
Then, due to \cref{eq:CP-Xi}, the $\Xi$ matrix takes the form
\beq
\Xi = \begin{pmatrix}
  0 &    \zeta \\
  \zeta^T & \Xi^\pr
\end{pmatrix} + \mcl{O}(\qq^2),
\eeq
where 
\beq
\zeta_{1,a} = \Xi_{1,a},\qquad a=2,3\cdots ,
\eeq
and
\beq
{\Xi}^\pr_{a,a^\pr} = \Xi_{a,a^\pr},\qquad a,a^\pr=2,3\cdots ,
\eeq
are submatrices of $\Xi$.
We emphasize that for $a\ge 2$, $\lambda_a$ is {\it not} singular.
Thus in the limit $\qq\to 0$, ${\Xi}^\pr$ approaches a constant matrix, whereas $\zeta_{1,a}\sim |\qq|$.
Therefore, by diagonalizing $\Xi^\pr$, we can rewrite $\Xi$ as
\beq
\Xi = U^T \begin{pmatrix}
  0        &  c_2|\qq|   & c_3|\qq|   & \cdots  \\
  c_2|\qq| &  \xi_2^\pr  & 0          & \cdots  \\
  c_3|\qq| &  0          & \xi^\pr_3  & \cdots  \\
  \vdots   &  \vdots     & \vdots     & \ddots
\end{pmatrix} U + \mcl{O}(\qq^2), \label{eq:CP-Xi}
\eeq
where $\xi^\pr_a$'s are eigenvalues of $\Xi^\pr$, and $U$ is some orthogonal matrix.
Now we prove that one of $\xi^\pr_{a=2,3\cdots}$ is zero.
We denote the diagonal matrix $\delta_{\nu,\nu^\pr} \chi_\nu$ as $[\chi]$.
Then the projected $[\chi]$ matrix in $Q$ subspace is $Q[\chi]Q=[\chi]-P[\chi]-[\chi]P$. 
Apparently, $\eta_\nu = 1$ and $\eta_\nu=\chi_\nu$ are two zero eigenvectors of $Q[\chi]Q$, wherein $\eta_\nu=1$ is in subspace $P$ whereas $\eta_\nu=\chi_\nu$ is in subspace $Q$. 
As $\Xi^\pr$ is equivalent to $Q[\chi]Q$ up to an invertable transformation, $\Xi^\pr$ has one zero eigenvalue in the Q subspace.
Therefore, one of $\xi^\pr_{a=2,3\cdots}$ is zero.
Here we choose $\xi^\pr_2=0$.
In the limit $\qq\to 0$, we have the $\Xi$ eigenvalues as
\beq
\xi_1(\qq) =-\xi_2(\qq) = |c_2| |\qq| + \mcl{O}(\qq^2),
\eeq
\beq
\xi_{n}(\qq) = \xi^\pr_n(\qq) +\mcl{O}(\qq^2),\qquad n=3\cdots N_\mrm{CP}(\BB) .
\eeq
Therefore, due to \cref{eq:omg_CP}, $n=1,2$ correspond to the gapped CP modes, whereas $n=3\cdots N_\mrm{CP}(\BB)$ correspond to the gapless CP modes.
The low energy behavior of the gapless CPs are very similar with the CZSs: both of them have a linear dispersion in the limit $\qq\to 0$.
However, a vital difference is that gapless CPs are coupled to gapped CPs, through the $c_{a\ge 3}$ terms in \cref{eq:CP-Xi}, whereas the CZSs cannot.
As a result, the dispersions of gapless and gapped CPs form anti-crossings, whereas dispersions of CZSs and gapped CPs form symmetry-protected crossings.

Here we give a simplified method to calculate the gapped CP frequency. 
Since the gapped CP is driven by the Coulomb interaction, for simplicity, in this method we omit $f_{\nu,\nu^\pr}$ and $1/\tau_\mrm{v}$.
From \cref{eq:Boltzmann-Son}, we get
\beq
\eta_\nu = \frac{\frac{e}{4\pi^2} (\qq\cdot\BB) }{\chi_\nu \beta_\nu(\BB) \omg- \frac{e}{4\pi^2} (\qq\cdot\BB)} \cdot \frac{e^2}{\ee_0\qq^2} \sum_{\nu^\pr} \eta_{\nu^\pr}, \label{eq:CP-n}
\eeq
and thus
\beq
1= \frac{e^2}{\ee_0\qq^2}  \sum_\nu \frac{\frac{e}{4\pi^2} (\qq\cdot\BB) \beta_\nu(\BB) }{\chi_\nu \beta_\nu(\BB) \omg  - \frac{e}{4\pi^2} (\qq\cdot\BB)}. \label{eq:CP-omg}
\eeq
Supposing $\omg$ is a constant in the limit $\qq\to 0$, we have
\beq
1 = \frac{e^2}{\ee_0\qq^2} \sum_\nu\pare{\chi_\nu \frac{\frac{e}{4\pi^2} (\qq\cdot\BB)}{\omg} 
+ \frac{\pare{\frac{e}{4\pi^2}\qq\cdot\BB}^2}{\beta_\nu(\BB) \omg^2} 
+ \chi_\nu \frac{\pare{\frac{e}{4\pi^2}\qq\cdot\BB}^3}{\beta_\nu^2(\BB) \omg^3} 
+ \frac{\pare{\frac{e}{4\pi^2}\qq\cdot\BB}^4}{\beta_\nu^3(\BB) \omg^4} 
+ \cdots}. \label{eq:CP-series} 
\eeq
To zeroth order of $\qq$, we need only keep first two terms in above equation. 
The first term must vanish due to the no-go theorem of Weyl semimetals \cite{Nielsen1981a,Nielsen1981b}, which says $\sum_\nu \chi_\nu = 0$.
Thus we have
\beq
\omg_{\mrm{CP},1,2} (\qq\to 0)= \pm \frac{e^2}{4\pi^2 } |\hqq\cdot\BB| \sqrt{\frac{1}{\ee_0} \sum_\nu \frac{1}{\beta_\nu(\BB)}}, \label{eq:omg_CP}
\eeq
where $\hqq = \qq/|\qq|$.


\section{Finite intra-valley scattering} \label{app:tau_0}
In this section, we solve \cref{eq:Boltzmann2} to first order of $\tau_0$ and justify the chiral limit approximation using a second order perturbation theory.
First, let us derive $\tau_0$ and $\tau_\mrm{v}$ explicitly in terms of scattering cross section.
We model the scattering cross section as
\beq
W_{\nu,\nu^\pr}(\sg,\sg^\pr) = \delta_{\nu,\nu^\pr}w_0 + (1-\delta_{\nu,\nu^\pr})w_1. \label{eq:W}
\eeq
Thus the scattering term is given by
\beqs
\SS[n_\nu(\sg)] = & \sum_{\nu^\pr} \int \frac{d^2\sg^\pr}{(2\pi)^3} \frac{\gm_{\nu^\pr}(\sg^\pr,\BB)}{|\vv_{\nu^\pr}(\sg^\pr)|} W_{\nu,\nu^\pr}(\sg,\sg^\pr) \pare{ \dn_{\nu^\pr}(\sg^\pr) - \dn_\nu(\sg) } \nono\\ 
= & (w_0-w_1) \int \frac{d^2\sg^\pr}{(2\pi)^3} \frac{\gm_{\nu}(\sg^\pr,\BB)}{|\vv_{\nu}(\sg^\pr)|} \Big[ \dn_\nu(\sg^\pr)-\dn_{\nu} (\sg) \Big]  +  w_1 \sum_{\nu^\pr} \int \frac{d^2\sg^\pr}{(2\pi)^3} \frac{\gm_{\nu^\pr}(\sg^\pr,\BB)}{|\vv_{\nu^\pr}(\sg^\pr)|} \Big[ \dn_{\nu^\pr} (\sg^\pr) - \dn_{\nu} (\sg) \Big]. \label{eq:scatering-tmp1}
\eeqs
We introduce the quasiparticle degree of freedom 
\beqs
\dbn_\nu &= \frac{1}{\beta_\nu(\BB)} \int \frac{d^2\sg}{(2\pi)^3} \frac{\gm_\nu(\sg,\BB)}{|\vv_\nu(\sg,\BB)|} \dn_\nu(\sg), \label{eq:dbn-def}
\eeqs
and $\dtn_\nu(\sg) = \dn_\nu(\sg) - \dbn_\nu$, then \cref{eq:scatering-tmp1} can be rewritten as
\begin{align}
\SS[n_\nu(\sg)] =& (w_0-w_1) \sbrak{\beta_\nu(\BB)\dbn_\nu - \beta_\nu(\BB)\dn_\nu(\sg)} + w_1 \sum_{\nu^\pr}\sbrak{ \beta_{\nu^\pr}(\BB) \dbn_{\nu^\pr} - \beta_{\nu^\pr}(\BB) \dn_{\nu}(\sg)} \nono\\
=& -(w_0-w_1) \beta_\nu(\BB) \dtn_\nu(\sg) + w_1 \sum_{\nu^\pr}\beta_{\nu^\pr}(\BB)(\dbn_{\nu^\pr}-\dbn_\nu) - w_1 \sum_{\nu^\pr} \beta_{\nu^\pr}(\BB) \dtn_\nu(\BB) \nono\\
=& -\pare{w_0\beta_\nu(\BB) + w_1\sum_{\nu^\pr\neq\nu} \beta_{\nu^\pr}(\BB)}\dbn_\nu(\sg) - \pare{w_1\sum_{\nu^\pr}\beta_{\nu^\pr}(\BB)} \dbn_\nu(\sg) + w_1\sum_{\nu^\pr} \beta_{\nu^\pr}(\BB) \dbn_{\nu^\pr}.
\end{align}
Defining
\beq
\frac{1}{\tau_{0,\nu}} = w_0 \beta_\nu(\BB) + w_1 \sum_{\nu^\pr \neq \nu} \beta_{\nu^\pr}(\BB), \label{eq:tau0}
\eeq
\beq
\frac{1}{\tau_{\mrm{v}}} = w_1 \sum_{\nu^\pr} \beta_{\nu^\pr},
\eeq
we can rewrite the scattering term as
\beq
\SS [\dn_\nu(\sg)] = -\frac{\dtn_\nu(\sg)}{\tau_{0,\nu}} - \frac{\dbn_\nu}{\tau_\mrm{v}} + w_1 \sum_{\nu^\pr} \beta_{\nu^\pr}(\BB) \dbn_{\nu^\pr}. \label{eq:scattering}
\eeq
The first term relaxes deformation of Fermi surfaces that does not change quasiparticle number in each valley, the second term relaxes the quasiparticle number in each valley, and the last term is a feedback from the change of total quasiparticle number.
Because the scattering term is elastic, the total quasiparticle number on the Fermi surface should be a constant under the scattering. 
One can confirm this by observing $\sum_\nu \int \frac{d^2\sg}{(2\pi)^3} \frac{\gm_\nu(\sg,\BB)}{|\vv_\nu(\sg,\BB)|} \mcl{S}[n_\nu(\sg)]= 0$.
For simplicity, in the following we will neglect the $\nu$-dependence in $\tau_{0,\nu}$, \ie setting $\tau_{0,\nu}=\tau_0$.

For simplicity, here we consider isotropic Fermi surfaces, where $|\vv_\nu(\sg)|$, $\beta_\nu(\sg)$, $\Omg_\nu(\sg)\cdot\hvv_\nu(\sg)$ do not depend on $\sg$, and $\gamma_\nu(\sg,\BB)=1$.
According to \cref{eq:Boltzmann2}, to leading order of $\tau_0$ we get the $\dtn_\nu(\sg)$ part as
\beq
\dtn_\nu(\sg) = -i \tau_0 (\qq\cdot\vv_\nu(\sg)) \frac{\pare{\omg+\frac{i}{\tau_\mrm{v}}} \chi_\nu}{\frac{e}{4\pi^2}(\qq\cdot\BB)} \eta_\nu + \mathcal{O}(\tau_0^2). \label{eq:eta1}
\eeq
Now we look at the leading order effect of $\tau_0$ on CZS modes.
For a specific branch of CZS, \cref{eq:eta1} gives
\beq
\dtn_\nu(\sg) = -i \tau_0 (\qq\cdot\vv_\nu(\sg)) \frac{\chi_\nu}{\ovl{\beta}(\BB) \xi} \eta_\nu + \mathcal{O}(\tau_0^2),
\eeq
where $\xi$ is the corresponding eigenvalue of $\Xi$ matrix (\cref{eq:CZS-Xi}).
Substiting it back into \cref{eq:Boltzmann2} and integrating $\sg$, we get
\beq
\pare{\omg+\frac{i}{\tau_\mrm{v}}}\chi_\nu \eta_\nu = -i\tau_0 \frac{\{ (\qq \cdot \vv_{F,\nu})^2\}}{\xi} \eta_\nu + \frac{e (\qq\cdot\BB)}{4\pi^2 \beta_\nu(\BB)}  \eta_\nu  + \frac{e}{4\pi^2} (\qq\cdot\BB) \sum_{\nu^\pr} \pare{ f_{\nu,\nu^\pr}  + \frac{e^2}{\ee_0\qq^2} } \eta_{\nu^\pr},
\eeq
where
\beq
\{ (\qq \cdot \vv_{F,\nu})^2\} = \frac{1}{\ovl{\beta}(\BB)} \int \frac{d^2\sg}{(2\pi)^3} \frac{1}{|\vv_\nu(\sg)|} (\qq\cdot\vv_{\nu}(\sg))^2
\eeq
Apparently, finite $\tau_0$ introduce a non-Hermitian term in the dynamic equation of $\eta$.
This term would lead to a damping rate proportional to $ \tau_0 q^2 v_F^2/\xi $.
Therefore, for zero sound to be stable, the following relation should be satisfied
\beq
\frac{eB q}{\beta \xi} \gg \tau_0 \frac{q^2 v_F^2}{\xi}. \label{eq:tau0-cond1}
\eeq
Considering the inter-valley scattering, the following relation should also be satisfied
\beq
\frac{eB q}{\beta \xi} \gg \frac{1}{\tau_\mrm{v}},
\eeq
The above two inequalities are equivalent to
\beq 
\frac{\xi}{\tau_v v_F}  \frac{\mu^2}{\omg_B^2 }
\ll q \ll 
\frac{1}{\tau_0 v_F}\frac{\omg_B^2}{\mu^2},
\eeq
which have solutions only if
\beq
\frac{\tau_0}{\tau_v} \ll \frac{\omg_B^4}{\mu^4} \frac{1}{\xi}. \label{eq:tau0-cond2}
\eeq
\cref{eq:tau0-cond2} gives the upper limit of $\tau_0$, above which the CZS modes become unstable.
It should be noticed that $1/\xi$ is in order of $1+\ovl{\beta}(\BB) |f|$.

Perturbation theory for gapless CP modes is similar with the perturbation theory for the CZS modes, and the stable condition of gapless CP modes is also given by \cref{eq:tau0-cond2}.
Now we consider the gapped CP modes.
For the positive branch of gapped CP, the frequency of which is denoted as $\omg_{\mrm{CP}}$, \cref{eq:eta1} gives
\beq
\dtn_\nu = -i \tau_0 (\hqq\cdot\vv_\nu(\sg)) \frac{\omg_\mrm{CP} \chi_\nu}{\frac{e}{4\pi^2}(\hqq\cdot\BB)} \eta_\nu + \mathcal{O}(\tau_0^2).
\eeq
Following the above analysis, we find this term would lead to a damping rate proportional to $\ovl{\beta}(\BB)  (q v_F) (\tau_0\omg_\mrm{CP}) /{eB} \sim (q v_F) (\tau_0 \omg_\mrm{CP}) \frac{\mu^2}{\omg_B^2}$.
Thus, for the gapped CP to be stabel, there should be
\beq
\tau_0 \ll \frac{1}{q v_F}\frac{\omg_B^2}{\mu^2}. \label{eq:tau0-cond3}
\eeq
Therefore, the CP modes in the longwave limit are always stable against the intra-valley scattering.

\section{$\kk$-dependent scattering and interaction} \label{app:k-dep}
In this section we consider the $\kk$-dependence in the scattering cross section and residual short-range interaction.
We will show that the dynamic equations in \cref{eq:Boltzmann-Son-text,eq:Boltzmann-Son} are still correct in the chiral limit except that the parameters should be modified.

\subsection{Elastic scattering conserving the renormalized energy}
We emphasize that it is the renormalized quasiparticle energy, other than the bare quasiparticle energy, that is conserved in the scattering process.
This effect is not considered in \cref{app:Boltzmann,app:Chiral,app:CP,app:CZS,app:tau_0}. 
As explained below, when the short-range interaction $f_{\nu,\nu^\pr}(\kk,\kk^\pr)$ is $\kk$-independent, this effect can be neglected safely. 
However, when $f_{\nu,\nu^\pr}(\kk,\kk^\pr)$ becomes $\kk$-dependent, it is crucial to consider this effect to obtain the correct dynamic equation.

In presence of a $\kk$-dependent interaction, the renormalized quasiparticle energy in \cref{eq:QE-q} is modified to be
\begin{equation}
\ee_\nu(\kk) =  \ee^0_\nu(\kk) + \sum_{ \nu^\pr}\int \frac{d^3\kk^\pr}{(2\pi)^3}  \pare{f_{\nu,\nu^\pr}(\kk,\kk^\pr) + \frac{e^2}{\ee_0\qq^2}} \gm_{\nu^\pr}(\kk^\pr,\BB)
\dn_{\nu^\pr}(\kk^\pr). \label{eq:QE-q-k}
\end{equation}
Now we neglect the $\rr$ and $t$ dependence in $\dn_\nu(\kk)$ because the scattering process has much shorter length scale and time scale than the collective mode. 
Here we have omitted the plane-wave factor $e^{i(\qq\cdot\rr-\omg t)}$ for simplicity.
Changing $\kk$ to the variable $\ee,\sg$ and writing $n_\nu(\kk)$ as $n_\nu(\ee,\sg) = n_F(\ee-\mu) + \delta(\ee-\mu) \dn_\nu(\sg)$ (as introduced in \cref{app:Boltzmann}), we can write the correcion to the quasiparticle energies of the quasiparticles on Fermi surface as
\begin{equation}
\ee^0_\nu(\kk)=\mu\qquad \Rightarrow \qquad \Delta_\nu(\sg) =  \ee_\nu(\kk) - \ee^0_\nu(\kk)= \sum_{ \nu^\pr}\int \frac{d^2\sg^\pr}{(2\pi)^3}  \pare{f_{\nu,\nu^\pr}(\sg,\sg^\pr) + \frac{e^2}{\ee_0\qq^2}} \frac{\gm_{\nu^\pr}(\kk^\pr,\BB)}{|\vv_{\nu^\pr}(\sg^\pr)|} \dn_\nu(\sg^\pr). \label{eq:Dsig-def}
\end{equation} 
We require the renormalized quasiparticle energy to be conserved in the scattering process. 
Thus scattering term is modified to be
\begin{equation}
\SS[n_\nu(\kk)] =  \sum_{\nu^\pr} \iint \frac{d\ee^\pr d^2\sg^\pr}{(2\pi)^3} \frac{\gm_{\nu}(\sg^\pr,\BB)}{|\vv_{\nu}(\sg^\pr)|} W_{\nu,\nu^\pr} (\sg,\sg^\pr) \delta\pare{\ee + \Delta_\nu(\sg) - \ee^\pr -\Delta_{\nu^\pr}(\sg^\pr)} \pare{ n_{\nu^\pr}(\ee^\pr,\sg^\pr) - n_\nu(\ee,\sg) }.
\end{equation}
To linear order of $\dn_\nu(\sg)$, we obtain
\begin{align}
\SS[n_\nu(\kk)] =&  \sum_{\nu^\pr} \iint \frac{d\ee^\pr d^2\sg^\pr}{(2\pi)^3} \frac{\gm_{\nu}(\sg^\pr,\BB)}{|\vv_{\nu}(\sg^\pr)|} W_{\nu,\nu^\pr} (\sg,\sg^\pr) \delta(\ee-\ee^\pr) \pare{ \delta(\ee^\pr-\mu) \dn_{\nu^\pr}(\sg^\pr) - \delta(\ee-\mu) \dn_\nu(\sg) } \nono\\
& +  \sum_{\nu^\pr} \iint \frac{d\ee^\pr d^2\sg^\pr}{(2\pi)^3} \frac{\gm_{\nu}(\sg^\pr,\BB)}{|\vv_{\nu}(\sg^\pr)|} W_{\nu,\nu^\pr} (\sg,\sg^\pr) \pare{ n_F(\ee + \Delta_\nu(\sg) - \Delta_{\nu^\pr}(\sg^\pr)-\mu) - n_F(\ee-\mu) } \nono\\
= & \delta(\ee-\mu) \sum_{\nu^\pr} \int \frac{d^2\sg^\pr}{(2\pi)^3} \frac{\gm_{\nu}(\sg^\pr,\BB)}{|\vv_{\nu}(\sg^\pr)|} W_{\nu,\nu^\pr} (\sg,\sg^\pr) \pare{ \dn_{\nu^\pr}(\sg^\pr) + \Delta_{\nu^\pr}(\sg^\pr) - \dn_\nu(\sg) -\Delta_{\nu}(\sg)}. 
\end{align}

\subsection{The valley degree of freedom}
In \cref{app:Chiral} we decomposed $\dn_\nu(\sg)$ into two parts: the valley degree of freedom $\dbn_\nu(\sg)$ and the Fermi surface degree of freedom $\dtn_\nu(\sg)$.
In \cref{app:tau_0} we showed that if the short-range interaction and the scattering cross section are $\kk$-independent, $\dbn_\nu(\sg)$ a constant for each valley (\cref{eq:dbn-def}).
In the following we will show that with $\kk$-dependent interaction and scattering the valley degree of freedom is still well defined but its form will be modified.

First, we decompose the scattering cross section into an intra-valley components and an inter-valley component
\begin{equation}
W_{\nu,\nu^\pr}(\sg,\sg^\pr) = \delta_{\nu,\nu^\pr} W^{(0)}_{\nu}(\sg,\sg^\pr) + (1-\delta_{\nu,\nu^\pr}) W^{(1)}_{\nu\nu^\pr}(\sg,\sg^\pr). \label{eq:W-k}
\end{equation}
Correspondingly, the we decompose the scattering term into an intra-valley term $\mcl{S}^{(0)}$ and an inter-valley term $\mcl{S}^{(1)}$.
Here we are only interested in $\mcl{S}^{(0)}$
\begin{equation}
\SS^{(0)}[\dn_\nu(\sg)] = \frac{1}{\lambda} \int \frac{d^2\sg^\pr}{(2\pi)^3} \frac{\gm_{\nu}(\sg^\pr,\BB)}{|\vv_{\nu}(\sg^\pr)|} W_{\nu}^{(0)} (\sg,\sg^\pr) \pare{ \dn_{\nu^\pr}(\sg^\pr) + \Delta_{\nu^\pr}(\sg^\pr) - \dn_\nu(\sg) -\Delta_{\nu}(\sg)}. \label{eq:SS0}
\end{equation}
The valley degrees of freedom are undamped under the intra-valley scattering.
The following condition is sufficient and necessary for $\dbn_\nu(\sg)$ to be undamped under arbitrary intra-valley scattering
\begin{equation}
\dbn_\nu(\sg) + \Delta_{\nu}(\sg) = \dbn_\nu(\sg) + \sum_{ \nu^\pr}\int \frac{d^2\sg^\pr}{(2\pi)^3}  \pare{f_{\nu,\nu^\pr}(\sg,\sg^\pr) + \frac{e^2}{\ee_0\qq^2}} \frac{\gm_{\nu^\pr}(\kk^\pr,\BB)}{|\vv_{\nu^\pr}(\sg^\pr)|} \dbn_{\nu^\pr}(\sg^\pr) = a_\nu, \label{eq:dbn-cond}
\end{equation}
where $a_\nu$ is some constant.
It is direct to see that when $f$ is independent of $\sg$ \cref{eq:dbn-def} satisfies \cref{eq:dbn-cond}.

We expand the valley degree of freedom on a set of basis functions 
\begin{equation}
\dbn_\nu(\sg) = \sum_\alpha c_\alpha h_{\nu \alpha}(\sg). \label{eq:dbn-k}
\end{equation}
For $\kk$-independent $f$ we can simply set $h_{\nu \alpha}(\sg) = \delta_{\nu \alpha}$ such that for arbitrary $c_\alpha$ \cref{eq:dbn-cond} is satisfied.
For $\kk$-dependent $f$ we require $h_{\nu\alpha}$ to satisfy
\begin{equation}
h_{\nu \alpha}(\sg) + \sum_{ \nu^\pr}\int \frac{d^2\sg^\pr}{(2\pi)^3}  \pare{f_{\nu,\nu^\pr}(\sg,\sg^\pr) + \frac{e^2}{\ee_0\qq^2}} \frac{\gm_{\nu^\pr}(\kk^\pr,\BB)}{|\vv_{\nu^\pr}(\sg^\pr)|} h_{\nu^\pr \alpha}(\sg^\pr) = A_{\nu\alpha}, \label{eq:h-cond}
\end{equation}
where $A_{\nu\alpha}$ is a matrix, such that for arbitrary $c_\alpha$ \cref{eq:dbn-cond} is satisfied and $a_\nu$ is given as 
\begin{equation}
a_\nu = \sum_\alpha c_\alpha A_{\nu \alpha}. \label{eq:a-c}
\end{equation}
We decompose the short-range interaction into a $\kk$-independent part $\ovl{f}$ and a $\kk$-dependent part $\delta f$ 
\begin{equation}
f_{\nu,\nu^\pr}(\sg,\sg^\pr) = \ovl{f}_{\nu,\nu^\pr} + \delta f_{\nu,\nu^\pr} (\sg,\sg^\pr). \label{eq:f-decompose}
\end{equation}
Then the basis functions subject to \cref{eq:h-cond} can be solved by series expansion in order of $\delta f$.
We take the trial solution
\begin{equation}
h_{\nu \alpha}(\sg) = \delta_{\nu \alpha} + \sum_{m=1}^\infty h^{(m)}_{\nu \alpha}, \label{eq:h-series}
\end{equation} 
where $h^{(m)}$ is in $m$-th order of $\delta f$.
Substiting \cref{eq:h-series} into \cref{eq:h-cond}, we obtain
\begin{equation}
h^{(m+1)}_{\nu \alpha}(\sg) = - \sum_{\nu^\pr} \int \frac{d^2\sg^\pr}{(2\pi)^3} \frac{\gm_\nu(\sg^\pr,\BB)}{|\vv_\nu(\sg^\pr)|}  \delta{f}_{\nu,\nu^\pr}(\sg,\sg^\pr) h^{(m)}_{\nu^\pr\alpha},\qquad m=0,1,2\cdots,\label{eq:h(m+1)}
\end{equation}
where $h^{(0)}_{\nu\alpha}=\delta_{\nu\alpha}$ and
\begin{equation}
A_{\nu \alpha} = \delta_{\nu \alpha} + \sum_{\nu^\pr} \int \frac{d^2\sg^\pr}{(2\pi)^3} \frac{\gm_\nu(\sg^\pr,\BB)}{|\vv_\nu(\sg^\pr)|}  \pare{ \ovl{f}_{\nu,\nu^\pr} + \frac{e^2}{\ee_0\qq^2}} \pare{\delta_{\nu^\pr\alpha} + \sum_{m=1}^\infty h^{(m)}_{\nu^\pr\alpha}(\sg^\pr)}.
\end{equation}
We can properly choose $\ovl{f}_{\nu,\nu^\pr}$ such that 
\begin{equation}\small
\int \frac{d^2\sg}{(2\pi)^3} \frac{\gm_\nu(\sg,\BB)}{|\vv_\nu(\sg)|} 
\int \frac{d^2\sg^\pr}{(2\pi)^3} \frac{\gm_\nu(\sg^\pr,\BB)}{|\vv_{\nu^\pr}(\sg^\pr)|} 
\pare{ \delta f_{\nu,\nu^\pr}(\sg,\sg^\pr) + \sum_{\nu^\prpr}\int \frac{d^2\sg^\prpr}{(2\pi)^3} \frac{\gm_\nu(\sg^\prpr,\BB)}{|\vv_{\nu^\prpr}(\sg^\prpr)|} \delta f_{\nu,\nu^\prpr}(\sg,\sg^\prpr) \delta f_{\nu^\prpr,\nu^\pr}(\sg^\prpr,\sg^\pr) +\cdots }= 0. \label{eq:df}
\end{equation}
Then due to \cref{eq:df,eq:h(m+1)} there are
\begin{equation}
\int \frac{d^2\sg}{(2\pi)^3} \frac{\gm_\nu(\sg,\BB)}{|\vv_\nu(\sg)|} h_{\nu\alpha}(\sg) = \delta_{\nu \alpha} \beta_\nu(\BB), \label{eq:h-int}
\end{equation}
and
\begin{equation}
A_{\nu\alpha} = \delta_{\nu \alpha} + \pare{\ovl{f}_{\nu,\alpha} + \frac{e^2}{\ee_0\qq^2}} \beta_{\alpha}(\BB). \label{eq:A-def}
\end{equation}
To be specific, we can expand $\ovl{f}_{\nu,\nu^\pr}$ that fulfills \cref{eq:df} in orders of $f$ as
\begin{equation}
\ovl{f} = \ovl{f}^{(1)} + \ovl{f}^{(2)} + \cdots,\label{eq:fbar}
\end{equation}
where 
\begin{equation}
\ovl{f}^{(1)}_{\nu,\nu^\pr} = \frac{1}{\beta_\nu(\BB)\beta_{\nu^\pr}(\BB)} 
\int \frac{d^2\sg}{(2\pi)^3} \frac{\gm_\nu(\sg,\BB)}{|\vv_\nu(\sg)|} 
\int \frac{d^2\sg^\pr}{(2\pi)^3} \frac{\gm_\nu(\sg^\pr,\BB)}{|\vv_{\nu^\pr}(\sg^\pr)|}
f_{\nu,\nu^\pr}(\sg,\sg^\pr), \label{eq:fbar1}
\end{equation}
and
\begin{align}\small
\ovl{f}^{(2)}_{\nu,\nu^\pr} =& \frac{1}{\beta_\nu(\BB)\beta_{\nu^\pr}(\BB)} 
\int \frac{d^2\sg}{(2\pi)^3} \frac{\gm_\nu(\sg,\BB)}{|\vv_\nu(\sg)|} 
\int \frac{d^2\sg^\pr}{(2\pi)^3} \frac{\gm_\nu(\sg^\pr,\BB)}{|\vv_{\nu^\pr}(\sg^\pr)|} \nono\\
& \times \pare{ \sum_{\nu^\prpr}\int \frac{d^2\sg^\prpr}{(2\pi)^3} \frac{\gm_\nu(\sg^\prpr,\BB)}{|\vv_{\nu^\prpr}(\sg^\prpr)|}  \pare{ f_{\nu,\nu^\prpr}(\sg,\sg^\prpr) - f^{(1)}_{\nu,\nu^\prpr} } \pare{ f_{\nu^\prpr,\nu^\pr}(\sg^\prpr,\sg^\pr) - f^{(1)}_{\nu^\prpr,\nu^\pr} } }.\label{eq:fbar2}
\end{align}

\subsection{dynamic equation}
Now we study the dynamic equation of the valley degrees of freedom.
We first look at the scattering term.
Since $dn_\nu(\sg)=\dbn_\nu(\sg) + \dtn_\nu(\sg)$ and $\SS=\SS^{(0)} + \SS^{(1)}$, where $\SS^{(0)}$ is contributed by intra-valley scattering (\cref{eq:SS0}) and $\SS^{(1)}$ is contribute by inter-valley scattering, the total scattering term decomposes into four terms 
\begin{equation}
\SS[n_\nu(\sg)] = \SS^{(0)}\sbrak{\dbn_\nu(\sg)} +  \SS^{(1)}\sbrak{\dbn_\nu(\sg)} + \SS^{(0)}\sbrak{\dtn_\nu(\sg)} + \SS^{(1)}\sbrak{\dtn_\nu(\sg)}.
\end{equation}
In last subsection we proved that $\SS^{(0)}\sbrak{\dbn_\nu(\sg)}=0$.
Now we make relaxation time approximation for the other three terms
\begin{equation}
\SS^{(1)}\sbrak{\dbn_\nu(\sg)} \approx -\frac{\dbn_\nu(\sg)}{\tau_\mrm{v}},
\end{equation}
\begin{equation}
\SS^{(0)}\sbrak{\dtn_\nu(\sg)} + \SS^{(1)}\sbrak{\dtn_\nu(\sg)} \approx -\frac{\dtn_\nu(\sg)}{\tau_0}.
\end{equation}
In the chiral limit we have $\SS^{(0)}\gg \SS^{(1)}$ and so $\tau_0\ll \tau_\mrm{v}$.

Following the derivation in \cref{app:Boltzmann}, we obtain the linearized Boltzmann's equation with $\kk$-dependent short-range interaction as
\begin{align}
&  \frac{\gm_\nu(\sg,\BB)}{|\vv_\nu(\sg)|} \pare{ \pare{\omg+\frac{i}{\tau_\mrm{v}}} \dbn_\nu(\sg)  + \pare{\omg+\frac{i}{\tau_\mrm{0}} }\dtn_\nu(\sg) } \nono\\
=&  \pare{\qq\cdot\hvv_\nu(\sg) - e\qq\cdot\BB(\hvv_\nu(\sg)\cdot\Omg_\nu(\sg))} \pare{\dn_\nu(\sg)+\Delta_\nu(\sg)}
- ie (\hvv_\nu(\sg) \x \BB)\cdot\pt_\kk \pare{\dn_\nu(\sg)+\Delta_\nu(\sg)}, \label{eq:Boltzmann1-k}
\end{align}
where $\Delta_\nu(\sg)$ is defined in \cref{eq:Dsig-def}.
To zeroth order of $\tau_0$, we have
\begin{equation}
  \dn_\nu(\sg) = \dbn_\nu(\sg) = \sum_\alpha c_\alpha h_{\nu \alpha}(\sg),
\end{equation} 
where $h_{\nu \alpha}(\sg)$ are the bases introduced in last subsection.
Due to \cref{eq:dbn-cond}, $\dn_\nu(\sg)+\Delta_\nu(\sg)$ is a constant $a_\nu$.
And due to \cref{eq:a-c,eq:A-def}, \cref{eq:Boltzmann1-k} can be written as
\begin{equation}
\frac{\gm_\nu(\sg,\BB)}{|\vv_\nu(\sg)|} \pare{\omg+\frac{i}{\tau_\mrm{v}}} \dbn_\nu(\sg)=  \pare{\qq\cdot\hvv_\nu(\sg) - e\qq\cdot\BB(\hvv_\nu(\sg)\cdot\Omg_\nu(\sg))} \pare{c_\nu + \sum_{\nu^\pr} \pare{ \ovl{f}_{\nu,\nu^\pr} + \frac{e^2}{\ee_0\qq^2}} \beta_{\alpha}(\BB) c_{\nu^\pr}}.
\end{equation}
Integrating $\sg$ on both sides of this equation and applying \cref{eq:h-int}, we obtain
\begin{equation}
\pare{\omg+\frac{i}{\tau_\mrm{v}}} \beta_\nu(\BB) c_\nu =  \chi_\nu \frac{e(\qq\cdot\BB)}{4\pi^2} \pare{c_\nu + \sum_{\nu^\pr}\pare{ \ovl{f}_{\nu,\nu^\pr} + \frac{e^2}{\ee_0\qq^2}} \beta_{\nu^\pr}(\BB) c_{\nu^\pr}}.
\end{equation}
We introduce the variable $\eta_\nu = \beta_\nu(\BB) c_\nu$ then we obtain
\begin{equation}
\pare{\omg+\frac{i}{\tau_\mrm{v}}} \chi_\nu \eta_\nu= \frac{e(\qq\cdot\BB)}{4\pi^2\beta_{\nu}(\BB)} \eta_\nu + \frac{e(\qq\cdot\BB)}{4\pi^2} \sum_{\nu^\pr}\pare{ \ovl{f}_{\nu,\nu^\pr} + \frac{e^2}{\ee_0\qq^2}}\eta_{\nu^\pr},
\end{equation}
which is of the same form of \cref{eq:Boltzmann-Son}.

\section{Thermaldynamic property of chiral zero sound} \label{app:thermal}
We treat the CZS modes as bosonic quasiparticle excitations.
For each branch of CZS modes $\omg_n(\qq)$ we assign a distribution function $g_n(\qq,\rr,t)$, and in equilibrium it's just the Bose-Einstein distribution, \ie 
\beq
g_n^{(0)}(\qq,\rr,t) = \frac{1}{ \exp\pare{\frac{\omg_n(\qq)}{k_\mrm{B}T}} - 1 },
\eeq
where $k_\mrm{B}$ is the Boltzmann's constant and $T$ is the temperature.
Here we have dropped the ``CZS'' subscript for brevity.
In the following we assume the magnetic field is applied along the $z$-direction, so the dispersion is $\omg_n(\qq) = c_n q_z $.

First let us calculate the specific heat per unit volume
\beqs
\kappa(T) & = \frac{\pt}{\pt T} \sum_n \int_{|\qq|<\Lambda} \frac{d^3 \qq}{(2\pi)^3}  \frac{|c_n q_z|}{\exp\pare{\frac{|c_n q_z|}{k_\mrm{B}T}}-1} \nono\\
& = \sum_n k_\mrm{B} \int_{-\Lambda}^{\Lambda} \frac{d q_z}{2\pi}  \frac{\Lambda^2-q_z^2}{4\pi} \pare{\frac{c_n q_z}{k_\mrm{B}T}}^2 \frac{1}{4\sinh^2 {\frac{c_n q_z}{2k_\mrm{B}T}}} \nono\\
& = \sum_n \kappa_{n}(T),
\eeqs
\beq
\kappa_{n}(T)=  k_\mrm{B} \Lambda^3 \pare{\frac{k_\mrm{B}T}{c_n\Lambda}}^3 \int_{-c_n\Lambda/(k_\mrm{B}T)}^{c_n\Lambda/(k_\mrm{B}T)} \frac{d x}{2\pi}  \frac{(c_n\Lambda/(k_\mrm{B}T))^2-x^2}{4\pi}  \frac{x^2}{4\sinh^2 {\frac{x}{2}}} ,
\eeq
where $\Lambda\sim 1/a_0$ is the cutoff of $q$, $a_0$ is lattice constant, and $\kappa_{n}$ is the specific heat contributed by the $n$-th branch of CZS.
In the two limits $c_n \Lambda \gg k_\mrm{B}T$ and  $c_n \Lambda \ll k_\mrm{B}T$,  we have
\beq
\kappa_{n}(T) = \begin{cases} \displaystyle
k_\mrm{B}\Lambda^3 \frac{k_\mrm{B}T}{12c_n\Lambda}, \qquad & c_n \Lambda \gg k_\mrm{B}T \\
\displaystyle k_\mrm{B}\Lambda^3\frac{1}{6\pi^2}, \qquad & c_n \Lambda \ll k_\mrm{B}T 
\end{cases}.
\eeq

Now let us calculate the thermal conductivity. 
For an inhomogeneous system, the distribution function satisfy the Boltzmann's equation
\beq
\pt_t g_n(\qq,\rr,t) = -  c_n \pt_z g_n(\qq,\rr,t) - \frac{g_n(\qq,\rr,t) - g_n^{(0)}(\qq,\rr,t)}{\tau_\mrm{s}(T)},
\eeq
where $\tau_\mrm{s}(T)$ is the relaxation time for the CZS excitations.
In low temperature, the relaxation should be proportional to $\tau_\mrm{v}$.
In presence of a temperature gradient, the first order stationary solution reads
\beq
\delta g_n(\qq,\rr) = g_n(\qq,\rr) - g_n^{(0)}(\qq,\rr) 
\approx - \tau_\mrm{s}(T) (c_n \pt_z T ) \frac{\omg_n(\qq)}{k_\mrm{B}T^2} \frac{1}{4\sinh^2\frac{\omg_n(\qq)}{2k_\mrm{B}T}}.
\eeq
The thermal current is given by
\beqs
j_z^\mrm{th} &= \sum_n \int_{|\qq|<\Lambda} \frac{d^3\qq}{(2\pi)^3} \omg_n(\qq) c_n \delta g_n(\qq,\rr) \nono\\
& = -  \tau_\mrm{s}(T) \sum_n \int_{|\qq|<\Lambda} \frac{d^3\qq}{(2\pi)^3} c_n ( c_n \pt_z T )  \frac{\omg_n^2(\qq)}{k_\mrm{B}T^2} \frac{1}{4\sinh^2\frac{\omg_n(\qq)}{2k_\mrm{B}T}}.
\eeqs
Therefore, the thermal conductivity is 
\beqs
\sigma^\mrm{th}_{i,j} = \delta_{i,z}\delta_{j,z} \tau_\mrm{s}(T) \sum_n c_n^2 \kappa_{n}(T).
\eeqs

\section{Strong magnetic field and finite temperature}\label{app:strongB}
The above derivations are based on Boltzmann's equation, which is valid only if $\omg_B \tau_0 \ll 1$, $\omg_B\ll \mu$.
Thus, it is still unknown whether CP and CZS modes exist in the case $\omg_B \tau_0 \gtrsim 1$, $\omg_B\ll \mu$.
Here we refer this case as the strong field case. 
In this case, the Landau levels are formed and there are many Landau levels under the the chemical potential.
Therefore, the system should be described by distribution functions on the Landau levels.
Here we expand this distribution function as a equilibrium part and a small deviation from equilibrium
\beq
n_\nu(k,\al,t) = n_T(\ee^0_\nu(k,\al)-\mu) + \delta_T (\ee^0_\nu (k,\al)-\mu) \dn_\nu(\al) e^{i(\qq\cdot\rr-\omg t)}.
\eeq
Here $k$ is the momentum along magnetic field, $\al$ is the Landau level index, $\ee^0_\nu(k,\al)$ is the Landau levels, $n_0(\ee^0_\nu(k,\al)-\mu)= \langle \psi_{k\al}^\dagger \psi_{k\al} \rangle$ is the occupation number in equilibrium, and $\delta_0(\ee) = -\pt_\ee n_0(\ee)$.
We assume the Landau levels as \cite{Abrikosov1998}
\beq
\ee^0_\nu(k,\al) = \begin{cases}
uk + v\sqrt{k^2 + 2eB \al}\qquad & \al>0 \\
uk + \chi v k\qquad &\al=0 \\
uk - v\sqrt{k^2 +2eB |\al|}\qquad & \al<0 
\end{cases}, \label{eq:LL}
\eeq
where $|u|<|v|$ such that the WP is type-I \cite{Soluyanov2015}.
In presence of scattering \cref{eq:W}, we can write the spectrum  function as \cite{Coleman2015}
\beq
A(\ee^0_\nu(\kk,\al),\omg) =\frac{1}{\pi} \frac{1/(2\tau_0)}{(\ee^0_\nu(\kk,\al)-\mu-\omg)^2 + 1/(2\tau_0)^2}
\eeq
where $\tau_0$ is the quasiparticle life time (\cref{eq:tau0}). 
Therefore, the occupation number is given by
\beqs
n_T(\ee)  = \int d\omg \frac{1}{1+\exp\frac{\omg}{k_\mrm{B}T}} A(\ee,\omg),
\eeqs
and its derivative is given by
\beq
\delta_T (\ee)= \int d\omg \frac{1/(2k_\mrm{B}T)}{1+\cosh\frac{\omg}{k_\mrm{B}T}} A(\ee,\omg). \label{eq:def-delta-T}
\eeq
Similar with the weak field case, we decompose $\dn_\nu(\al)$ as a valley degree
\beq
\dbn_\nu = \frac{1}{\beta_\nu(B)} \frac{eB}{2\pi}\int \frac{dk}{2\pi} \sum_\al  \delta_T (\ee^0_\nu (k,\al)-\mu) \dn_\nu(\al),
\eeq
and a Fermi surface degree
\beq
\dtn_\nu(\al) = \dn_\nu(\al) - \dbn_\nu,
\eeq
where 
\beq
\beta_\nu(B) = \frac{eB}{2\pi}\int \frac{dk}{2\pi} \sum_\al  \delta_T (\ee^0_\nu (k,\al)-\mu)
\eeq
is the compressibility at finite temperature.
Here we have assumed that $eB>0$.
Then the kinetic equation of collective modes can be written as
\beqs
& \delta_T(\ee_\nu^0(k,\al)-\mu) \pare{ \pare{\omg +\frac{i}{\tau_\mrm{v}}}\dbn_\nu + \pare{\omg +\frac{i}{\tau_0}}\dtn_\nu(\al) }  \nono \\
= &   \delta_T(\ee_\nu^0(k,\al)-\mu) 
\qq\cdot\vv_\nu(\al) \pare{\dn_\nu(\al) + \sum_{\nu^\pr} \pare{f_{\nu,\nu^\pr}+\frac{e^2}{\ee_0\qq^2}} \beta_{\nu^\pr}(B) \dbn_{\nu^\pr}}.
\eeqs
We define the inequilibrium quasiparticle number in the $\nu$-th valley as $\eta_\nu = \beta_\nu(B)\dbn_\nu$.
Then, to zeroth order of $\tau_0$, integrating $k$ and suming over $\al$, we get
\beq
\pare{\omg +\frac{i}{\tau_\mrm{v}}} \chi_\nu \eta_\nu = \frac{e(\qq\cdot\BB)}{4\pi^2 \beta_\nu(B)} \eta_\nu + 
\frac{e(\qq\cdot\BB)}{4\pi^2} \sum_{\nu^\pr} \pare{f_{\nu,\nu^\pr}+\frac{e^2}{\ee_0\qq^2}} \eta_{\nu^\pr},
\eeq
which has the exact same form with \cref{eq:Boltzmann-Son}. 
It should be noticed that, due to \cref{eq:LL}, only the zeroth Landau level contribute to the integral in the r.h.s..
One can easily verify that, the leading order effect of $\tau_\mrm{v}$ is introducing an effective damping rate, and, the stable condition for CZS and CP modes are still given by \cref{eq:tau0-cond2} and \cref{eq:tau0-cond3}, respectively.



\section{Quantum oscillation in compressibility} \label{app:dos-B}

After the Landau levels are formed, the compressibility at finite temperature is given by
\beq
\beta_\nu(B) = \sum_\al F(\al),
\qquad
F(\alpha) = \frac{eB}{2\pi} \int \frac{dk}{2\pi} \delta_0(\ee^0_\nu(k,\al)-\mu).
\eeq
Here we assume $\mu>0$.
Using the Poison's equation, we get \cite{Lifshitz2013}
\beqs
\beta_\nu(B) &= F(0)-\frac{1}{2}F(0^+) + \frac{1}{2} F (0^+) + \sum_{\al=1}^\infty F(\al)
\nono\\
& =F(0)-\frac{1}{2}F(0^+)+ \int_0^\infty d\al F(\al) + 2\Re \sum_{l=1}^\infty \int_0^\infty d\al F(\al) e^{2\pi i l \al}.
\eeqs
We define 
\beq
\beta_\nu^\odz(B) = F(0) - \frac{1}{2} F(0^+) + \int_0^\infty d\al F(\al) 
\eeq
as the not oscillating component, and 
\beq
\beta_\nu^{(l)}(B) = 2\Re \int_0^\infty d\al F(\al) e^{2\pi i l \al}, \qquad l\ge 1
\eeq
as the oscillating components.
$\beta^\odz_\nu(B)$ is just the compressibility in weak field limit (\cref{eq:rho-B}).
Now let us calculate $\beta^{(l)}_\nu(B)$.
Using the relation
\beq
\alpha = \frac{S_\nu(\ee,k)}{2\pi eB} - \phi_\nu(\ee,k),
\eeq
where $S_\nu$ is the area enclosed by the fixed-energy circle in the $k$-plane. $\phi(\ee,k)$ includes Maslov's index plus Berry's phase.
For linear isotropic WPs we always have $\phi_\nu(\ee,\kk)=0$, and so in the following we omit $\phi_\nu(\ee,k)$. 
Expanding $S_\nu(\ee,k)$ as
\beq
S_\nu(\ee,k) \approx S_{\mrm{ex},\nu}(\ee) + \frac{1}{2}\frac{\pt^2 S_\nu}{\pt k^2} \Big|_\mrm{ex}  (k-k_\mrm{ex})^2
\eeq
Then we have 
\beqs
\beta^{(l)}_\nu(B) &= \Re \frac{eB}{\pi } \int_0^\infty d\al \int \frac{dk}{2\pi}
\exp\pare{i2\pi l \al}  \delta_T (\ee^0_\nu(k,\al)-\mu) \nono\\
& = \Re \frac{1}{2\pi^2 } \int d\ee \int \frac{dk}{2\pi} \frac{dS_{\mrm{ex},\nu}(\ee)}{d\ee}
\exp\pare{i2\pi l \frac{S_{\mrm{ex},\nu}(\ee) + \frac{1}{2}\frac{\pt^2 S_\nu}{\pt k^2} \Big|_\mrm{ex}  (k-k_\mrm{ex})^2}{2\pi eB}}   \delta_T (\ee-\mu)
\eeqs
Applying the Gaussian integral formula $ \int_{-\infty}^\infty \exp\pare{\frac{i}{2}ax^2} dx = \pare{\frac{2\pi i}{a}}^\frac12 $, we get
\beqs
\beta^{(l)}_\nu(B) = \Re \frac{1}{4\pi^3 }  \int d\ee \pare{l \abs{ \frac{\pt^2 S_\nu}{\pt k^2}}_\mrm{ex} \frac{1}{2\pi eB}}^{-\frac12}   \frac{dS_{\mrm{ex},\nu}(\ee)}{d\ee} \exp\pare{i2\pi l \frac{S_{\mrm{ex},\nu}(\ee)}{2\pi eB} - i\frac{\pi}{4}} \delta_T (\ee-\mu),
\eeqs
where we assume $\frac{\pt^2 S_\nu}{\pt k^2}<0$.
Substiting \cref{eq:def-delta-T} into the above equation, we get
\beqs
\beta^{(l)}_\nu(B) = &  \Re  \frac{1}{4\pi^3} \int d\ee \int d\omg \pare{l \abs{ \frac{\pt^2 S_\nu}{\pt k^2}}_{\mrm{ex}}\frac{1}{2\pi eB}}^{-\frac12} \frac{dS_{\mrm{ex},\nu}(\ee)}{d\ee} \exp\pare{i2\pi l \frac{S_{\mrm{ex},\nu}(\ee)}{2\pi eB} - i\frac{\pi}{4}} \nono\\
& \times \frac{1/(2k_\mrm{B}T)}{1+\cosh(\frac{\omg}{k_\mrm{B}T})} \frac1{\pi} \frac{1/{2\tau_0}}{(\ee-\mu-\omg)^2+1/(2\tau_0)^2}.
\eeqs
Using contour integral in the upper half-plane of $\ee$, we get
\beqs
\beta^{(l)}_\nu(B) = &  \Re  \frac{1}{4\pi^3} \int d\omg \pare{l \abs{ \frac{\pt^2 S_\nu}{\pt k^2}}_{\mrm{ex}}\frac{1}{2\pi eB}}^{-\frac12} \frac{dS_{\mrm{ex},\nu}(\ee)}{d\ee}\bigg|_{\ee=\mu+\omg+i/(2\tau_0)} \exp\pare{i2\pi l \frac{S_{\mrm{ex},\nu}(\mu+\omg+\frac{i}{2\tau_0})}{2\pi eB} - i\frac{\pi}{4}} \nono\\
& \times \frac{1/(2k_\mrm{B}T)}{1+\cosh\frac{\omg}{k_\mrm{B}T}}.
\eeqs
We approximate $S_{\mrm{ex},\nu}(\mu+\omg+\frac{i}{2\tau_0})$ as $S_{\mrm{ex},\nu}(\mu) + \frac{d S_{\mrm{ex},\nu}(\mu)}{d\mu} \pare{\omg+\frac{i}{2\tau_0}} $, then we have
\beqs
\beta^{(l)}_\nu(B) \approx & \Re \frac{1}{8\pi^3}  \frac{dS_{\mrm{ex},\nu}(\mu)}{d\mu}  \pare{l \abs{ \frac{\pt^2 S_\nu}{\pt k^2}}_{\mrm{ex}}\frac{1}{2\pi eB}}^{-\frac12}\exp\pare{- \frac{dS_{\mrm{ex},\nu}(\mu)}{d\mu} \frac{ l}{2eB\tau_0} } \exp\pare{i2\pi l \frac{S_{\mrm{ex},\nu}(\mu)}{2\pi eB} - i\frac{\pi}{4}} \nono\\
&  \times \int d\omg \frac{1}{k_\mrm{B}T} \frac{\exp\pare{ i\frac{dS_{\mrm{ex},\nu}(\mu)}{d\mu} \frac{2 \omg l}{2eB} } }{1+\cosh\frac{\omg}{k_\mrm{B}T}}
\eeqs
Now we need to calculate the integral in second line of the above equation. 
We denote this integral as $I$.
We choose the contour $-\infty \to \infty \to \infty + i2\pi \to -\infty + i2\pi \to -\infty$, then we have
\beq
\pare{ 1-\exp\pare{-2\pi \frac{dS_{\mrm{ex},\nu}(\mu)}{d\mu} \frac{ k_\mrm{B}T l}{eB}} } I = 4 \pi \frac{dS_{\mrm{ex},\nu}(\mu)}{d\mu} \frac{ k_\mrm{B}T l}{eB}\exp\pare{ -\pi \frac{dS_{\mrm{ex},\nu}(\mu)}{d\mu} \frac{ k_\mrm{B}T l}{eB} }, 
\eeq
and thus
\beq
I = \frac{2}{\mrm{sinch}\pare{ \frac{dS_{\mrm{ex},\nu}(\mu)}{d\mu} \frac{ \pi k_\mrm{B}T l}{eB} }},
\eeq
where $\mrm{sinch}(x) = (e^x-e^{-x})/(2x)$.
Therefore, we get
\beqs
\beta^{(l)}_\nu(B) \approx & \frac{1}{4\pi^3}  \frac{dS_{\mrm{ex},\nu}(\mu)}{d\mu}  \pare{l \abs{ \frac{\pt^2 S_\nu}{\pt k^2}}_{\mrm{ex},\mu}\frac{1}{2\pi eB}}^{-\frac12}
\frac{ \exp\pare{- \frac{dS_{\mrm{ex},\nu}(\mu)}{d\mu} \frac{ l}{2eB\tau_0} }  }{ \mrm{sinch}\pare{  \frac{dS_{\mrm{ex},\nu}(\mu)}{d\mu} \frac{\pi k_\mrm{B}T l}{eB}  } }
\cos\pare{2\pi l \frac{S_{\mrm{ex},\nu}(\mu)}{2\pi eB} - \frac{\pi}{4}}.
\eeqs 

For an isotropic Fermi surface, where $\ee = v_F |\kk|$ and $S(\ee,k) = \pi \pare{ (\ee/v_F)^2-k^2}$, the first order oscillation is given by
\beq
\beta^{(1)}_\nu(B) \approx \frac1{2\pi^2} \frac{\mu \omg_B }{ v_F^3} 
\frac{ \exp\pare{ - 2\pi \frac{\mu }{\omg_B^2 2\tau_0}  } }{ \mrm{sinch}\pare{2\pi^2\frac{\mu k_\mrm{B}T}{\omg_B^2}} }
\cos\pare{\frac{S_{\mrm{ex},\nu}(\mu)}{eB} - \frac{\pi}{4}},
\eeq
and
\beq
\frac{\beta^\odf_\nu(B)}{\beta^\odz_\nu(B)} 
\approx \frac{\omg_B}{\mu} \frac{ \exp\pare{ - \pi \frac{\mu }{\omg_B^2 \tau_0}  } }{ \mrm{sinch}\pare{2\pi^2\frac{\mu k_\mrm{B}T}{\omg_B^2}} }
\cos\pare{\frac{S_{\mrm{ex},\nu}(\mu)}{eB} - \frac{\pi}{4}}.
\eeq

\end{document}